\documentclass[numberedappendix]{emulateapj}
\usepackage{apjfonts}
\usepackage{subfigure}
\newcommand{\detg}{{\sqrt{-g}}}
\newcommand{\del}{{\partial}}
\newcommand{\KS}{{\rm KS}}
\newcommand{\BL}{{\rm BL}}

\newcommand{\msun}{{\rm M_{\odot}}}

\newcommand{\dF}{{^{^*}\!\!F}}

\newcommand{\yr}{{\rm\,yr}}

\newcommand{\cm}{{\rm\,cm}}
\newcommand{\erg}{{\rm\,erg}}
\newcommand{\ergps}{{\rm\,erg~s^{-1}}}

\newcommand{\IEDEN}{{u_g}}

\shortauthors{McKinney, J.C.}
\shorttitle{Black Hole Jet Formation: Theory}
\begin{document}
\submitted{June 16, 2005}

\journalinfo{Web link for High Resolution document in footnote}

\title{Jet Formation in Black Hole Accretion Systems I: Theoretical
  Unification Model }
\author{Jonathan C. McKinney$^{1}$}
\altaffiltext{1}{Institute for Theory and Computation,
  Harvard-Smithsonian Center for Astrophysics, 60 Garden Street, MS
  51, Cambridge, MA 02138, USA\\
  High Res. Figures:
  \url{http://rainman.astro.uiuc.edu/\textasciitilde
    jon/j1.pdf}}
\email{jmckinney@cfa.harvard.edu}

\begin{abstract}

Two types of relativistic jets are suggested to form near accreting
black holes: a potentially ultrarelativistic Poynting-dominated jet
and a Poynting-baryon jet. One source of jet matter is
electron-positron pair production, which is driven by neutrino
annihilation in GRBs and photon annihilation in AGN and x-ray
binaries.  GRB Poynting-dominated jets are also loaded by
electron-proton pairs by the collisional cascade of Fick-diffused free
neutrons. We show that, for the collapsar model, the neutrino-driven
enthalpy flux (classic fireball model) is probably dominated by the
Blandford-Znajek energy flux, which predicts a jet Lorentz factor of
$\Gamma\sim 100-1000$. We show that radiatively inefficient AGN, such
as M87, are synchrotron-cooling limited to $\Gamma\sim
2-10$. Radiatively efficient x-ray binaries, such as GRS1915+105, are
Compton-drag limited to $\Gamma \lesssim 2$, but the jet may be
destroyed by Compton drag. However, the Poynting-baryon jet is a
collimated outflow with $\Gamma \sim 1-3$. The jet from radiatively
efficient systems, such as microquasar GRS1915+105, may instead be a
Poynting-baryon jet that is only relativistic when the disk is
geometrically thick.  In a companion paper, general relativistic
hydromagnetic simulations of black hole accretion with pair creation
are used to simulate jet formation in GRBs, AGN, and x-ray binaries.

\end{abstract}


\maketitle

\section{Introduction}\label{introduction}

Jets are a common outcome of accretion, yet the observed jet
properties, such as collimation and speed, are not uniform between
systems.  This is despite the fact that the basic physics (general
relativistic magnetohydrodynamics (GRMHD)) to describe such systems is
black hole mass-invariant.  Thus, it is worth-while to determine the
unifying, or minimum number of, pieces of physics that would explain
most of the features of gamma-ray bursts (GRBs), x-ray binaries, and
active galactic nuclei (AGN) \citep{gc02,ghis03,meier2003}.  To
understand jet {\it formation} requires at least explaining the origin
of the energy, composition, collimation, and Lorentz factor.  The goal
of this paper, and the companion numerical models paper
\citep{mckinney2005b}, is to explain these for GRBs, AGN, and x-ray
binaries.

Primarily we discuss two types of jets: Poynting-dominated jets
typically dominated in energy flux by Poynting flux and dominated in
mass by electron-positron pairs for AGN and x-ray binaries, while
dominated in mass by electron-proton pairs for GRBs ; and
Poynting-baryon jets with about equal Poynting flux and rest-mass flux
and dominated in mass by baryons.  The latter are sometimes referred
to as coronal outflows due to their origin. Generically this model is
similar to, e.g., \citet{sol89}, while here the motivation is based
upon the results of recent GRMHD numerical models.  This two-component
jet model is one key to understanding the diversity of jet
observations.  The Poynting-dominated jet is likely powered by the
Blandford-Znajek effect, while the Poynting-baryon jet is likely
powered by both Blandford-Znajek power and the release of disk
gravitational binding energy \citep{mckinney2005a}.  Collimation of
the polar Poynting-dominated jet may be due transfield balance against
the broader Poynting-baryon jet or by self-collimating hoop stresses.

Among all the black hole accretion systems, it appears that the least
unifiable is the observed emission.  While the radiative physics is
not black hole mass invariant, the observed differences suggest that
the environment likely plays a significant role in the emission.  For
example, while both blazars and GRBs exhibit non-thermal emission,
long-duration GRBs are harder with higher luminosity, while blazars
are softer with higher luminosity \citep{ghirlanda2004,ghirlanda2005}.
Also, GRBs lead to apparently most of the energy in $\gamma$-rays and
less than $10\%$ to the sub-$\gamma$-ray afterglow \citep{piran2005}.
On the contrary, blazars apparently release only $10\%$ in
$\gamma$-rays and the rest is produced in the radio lobe \citep{gc02}.
Despite the difficulties in understanding the emission processes in
some jet systems, the jet itself is probably produced by a universal
process.

The disk {\it and} jet radiative physics are keys to understanding the
evolution of the jet and why different systems have different terminal
Lorentz factors.  Through radiative annihilation of photons in AGN and
x-ray binary systems, the radiative physics may illuminate the origin
of jet composition by determining the electron-positron mass-loading
the Poynting-lepton jet, and so the Lorentz factor of the jet.  For
GRBs, the radiative annihilation of neutrinos and the effect of Fick
diffusion by free neutrons from the corona into the jet \citep{le2003}
may give an understanding of the Lorentz factor of the jet and the
origin of baryon-contamination.

The rest of this section briefly reviews the types of black hole
accretion systems and discusses jets in each.  At the end is an
outline of the paper.

\subsection{GRBs}

Neutron stars and black holes are associated with the most violent of
post-Big Bang events: supernovae and some gamma-ray bursts (GRBs) and
probably some x-ray flashes (XRFs) (for a general review see
\citealt{w93,wyhw00}).  Observations of a supernova light curve
(SN2003dh) in the afterglow of GRB 030329 suggest that at least some
long-duration GRBs are probably associated with core-collapse events
\citep{stanek2003,kawabata2003,uemura2003, hjorth2003}.

Neutrino processes and magnetic fields are both important to
understand core-collapse.  In unraveling the mechanism by which
core-collapse supernovae explode, the implementation of accurate
neutrino transport has been realized to be critical to whether a
supernova is produced in simulations \citep{Trans}. This has thus far
been interpreted to imply that highly accurate neutrino transport
physics is required, but this could also mean additional physics, such
as a magnetic field, could play a significant role. Indeed, all
core-collapse events may be powered by MHD processes rather than
neutrino processes \citep{lw70,sym84,ww86,dt92,khok99,awml03}.
Core-collapse involves shearing subject to the Balbus-Hawley
instability as in accretion disks \citep{awml03}.  All core-collapse
explosions are significantly polarised, asymmetric, and often bi-polar
indicating a strong role of rotation and a magnetic field (see, e.g.,
\citealt{ww96,wyhw00,wang01,wang02,wang03}, and references
therein). Possible evidence for a magnetic dominated outflow has been
found in GRB 021206 \citep{cb03}, marginally consistent with a
magnetic outflow directly from the inner engine \citep{lpb03},
although these observations remain controversial.

Black hole accretion is the key source of energy for many GRB models.
Collapsar type models suggest that a black hole forms during the
core-collapse of some relatively rapidly rotating massive stars.  The
typical radius of the accretion disk likely determines the duration of
long-duration GRBs \citep{w93,pac98,mw99}.  An accretion disk is also
formed as a result of a neutron star or black hole collisions with
another stellar object \citep{narayan1992,narayan2001}.

GRBs are believed to be the result of an ultrarelativistic
jet. Indirect observational evidence of relativistic motion is
suggested by afterglow achromatic light breaks and the ``compactness
problem'' suggests GRB material must be ultrarelativistic with Lorentz
factor $\Gamma\gtrsim 100$ to emit the observed nonthermal
$\gamma$-rays (see, e.g., \citealt{piran2005}).  Direct observational
evidence for relativistic motion comes from radio scintillation of the
ISM \citep{goodman1997} and measurements of the afterglow emitting
region from GRB030329 \citep{taylor2004a,taylor2004b}.

Typical GRB jet models invoke either a hot neutrino-driven jet or a
cold Poynting flux-dominated jet, while both allow for comparable
amounts of the accretion energy to power the jet \citep{pwf99}.  A
neutrino-driven jet derives its energy from neutrino annihilation from
gravitational energy and the jet is thermally accelerated.  However,
strong outflows can be magnetically driven
\citep{br76,lovelace76,blandford76}.  In particular, black hole
rotational energy can be extracted as a Poynting outflow \citep{bz77}.

\subsection{X-ray Binaries}

Long after their formation, neutron stars and black holes often
continue to produce outflows and jets \citep{mr99}.  These
include x-ray binaries (for a review see
\citealt{lewin1995,mcclintock2003}), neutron star as pulsars (for a
review see \citealt{lorimer2001} on ms pulsars and \citealt{tc99} on
radio pulsars) and soft-gamma ray repeaters (SGRs)
\citep{td95,td96,kouv1999}.  In the case of x-ray binaries, the
companion star's solar-wind or Roche-lobe forms an accretion disk.
Many x-ray binaries in their hard/low state (and radio-loud AGN) show
a correlation between the x-ray luminosity and radio luminosity
\citep{merloni2003}, which is consistent with radio synchrotron
emission from a jet and x-ray emission from a geometrically thick,
optically thin, Comptonizing disk.

Some black hole x-ray binaries have jets
\citep{mirabel1992,fender2003}, such as GRS 1915+105 with apparently
superluminal motion ($\Gamma\sim 3$) \citep{mr94,mr99,fb04}, but may
have $\Gamma\sim 1.5$ \citep{kaiser04}. Synchrotron radiation from the
jet suggests the presence of a magnetized accretion disk. Observations
of a broad, shifted, and asymmetric iron line from GRS 1915+105 is
possible evidence for a relativistic accretion disk
\citep{martocchia02}, although this feature could be produced by a jet
component.

The standard paradigm is that relativistic jets from x-ray binaries
are probably produced by the Blandford-Znajek effect.  However,
\citet{gd04} suggest that at least some black holes, such as GRS
1915+105, have slowly rotating black holes.  If this is correct, then
another mechanism is required to produce jets.  Indeed, jets or
outflows are produced from systems containing NSs, young stellar
objects, supersoft x-ray white dwarfs, symbiotic white dwarfs, and
even UV line-driven outflows from massive O stars.  Indeed, a
baryon-loaded coronal outflow with $\Gamma\sim 1.5-3$ can be produced
from a black hole accretion disk and not require a rapidly rotating
black hole \citep{mg04}.  Nonrelativistic outflows were found even in
viscous hydrodynamic simulations \citep{spb99,ia99,ia00,mg02}.  Such
baryon-loaded outflows or jets are sufficient to explain most known
x-ray binaries without invoking rapidly rotating black holes, and thus
unifies such mildly relativistic jets in neutron star and black hole
x-ray binaries.

\subsection{AGN}

Active galactic nuclei (AGN) have long been believed to be powered by
accretion onto supermassive black holes \citep{z64,salpeter64}.
Observations of MCG 6-30-15 show an iron line feature consistent with
emission from a relativistic disk with $v/c\sim 0.2$
\citep{tanaka1995, fabianvaughan2002}, although the lack of a temporal
correlation between the continuum emission and iron-line emission may
suggest it is a jet-related feature \citep{elvis2000}.

AGN are observed to have jets with $\Gamma\lesssim 10$
\citep{up95,biretta99}, even $\Gamma\sim 30$
\citep{brs94,gc01,jorstad01}, while some observations imply
$\Gamma\lesssim 200$ \citep{ghis93,kraw02,kono03}.  Some radio-quiet
AGN show evidence of weak jets \citep{ghis04}, which could be
explained as a coronal outflow \citep{mg04} and not require a rapidly
rotating black hole.  Observations imply the existence of a
two-component jet structure with a Poynting jet core and a dissipative
surrounding component \citep{ghis96,ghis05}.  The energy structure of
the jet and wind are important in understanding the feedback effect
that controls size of the black hole and may determine the $M-\sigma$
relation \citep{springel2004,dimat05}.

\subsection{Outline of Paper}

\S~\ref{modelsum} summarizes the proposed unified model to explain
jet formation in all black hole accretion systems.

\S~\ref{nonidealmhd} discusses why ideal MHD must break down in
magnetospheres and why the Goldreich-Julian charge density is never
reached.  A preliminary model is derived that describes the
pair-loading and baryon-loading of the Poynting-dominated jet.

\S~\ref{lorentzfactor} determines the Lorentz factor of
Poynting-dominated jets. The GRB jet Lorentz factor is shown to be
based upon electron-positron pair and baryon loading.  We show that
GRBs likely have electron-proton jets with Lorentz factor at large
distances of $100\lesssim\Gamma_{\infty}\lesssim 10^3$.

Based upon pair creation rates for AGN and x-ray binaries, we show
that relativity low radiatively efficient AGN, such as M87, have
electron-positron jets with $2\lesssim \Gamma_\infty\lesssim 10$.
Radiatively efficient systems, such as microquasar GRS1915+105, likely
do not have Poynting-dominated lepton jets but rather the observed
jets are a relativistic coronal outflow from the inner-disk.

\S~\ref{mixed} discusses Poynting-baryon jets and how they can explain
various observational features of jets in AGN and x-ray binaries.

\S~\ref{sumnumerical} summarizes the key results and fits from
GRMHD numerical models \citep{mckinney2005b} used in this paper.

\S~\ref{discussion} discusses the results and their possible
implications.

\S~\ref{conclusions} summarizes the key points.

Appendix~\ref{GRMHD} discusses breakdown of ideal MHD by
electron-positron pair creation by radiative annihilation and
electron-proton pair creation by ambipolar and Fick diffusion.  See
also the discussion in \citet{mckinney2005b}.
Appendix~\ref{conservedflow} gives a succinct summary of conserved
flow quantities in GRMHD used in section~\ref{lorentzfactor}.
Appendix~\ref{fluidforces} gives a derivation for the lab frame
stationary GRMHD forces along and perpendicular to the flow (field)
line in the lab frame.  This elucidates the origin of acceleration and
collimation.  Appendix~\ref{compton} gives the formulae for
Comptonization and pair annihilation used in
section~\ref{lorentzfactor}.

\section{GRMHD Pair Injection Model of Jet Formation}\label{modelsum}

The jet energy, composition, collimation, and Lorentz factor are
likely determined in a similar way for all black hole accretion
systems.  The particle acceleration mechanism and particle composition
of the jet remained unexplained in \citet{mg04}.  However, if field
lines tie the black hole to large distances, then the source of matter
is likely pair creation since the amount of matter that diffuses
across field lines is much smaller \citep{phi83,le93,punsly2001}.
Thus, the Poynting-dominated jet composition is electron-positron pair
dominated in AGN and x-ray binaries.

However, in GRB systems, free neutrons lead to baryon contamination
due to Fick diffusion across the field lines and subsequent rapid
collisionally-induced avalanche decay to an electron-proton plasma
\citep{le2003}.  The pair annihilation rates are much faster than the
dynamical time, and due to the temperature decrease, the
electron-positron pair rest-mass exponentially drops beyond the
fireball formation near the black hole.  Thus, the GRB jet composition
is likely dominated by electron-proton pairs.

GRMHD numerical models confirmed that accretion of a thick disk with
height ($H$) to radius ($R$) ratio of $H/R\gtrsim 0.1$ with a
homogeneous poloidal orientation self-consistently creates large scale
fields that tie the black hole to large distances
\citep{mg04,hirose04}.  Accretion of an irregular field loads the jet
with baryons and lowers the speed of the jet.  However, the existence
of a mostly uniform field threading the disk arises naturally during
core-collapse supernovae and NS-BH collision debris disks.  In AGN and
solar-wind capture x-ray binary systems, the accreted field is
probably uniform \citep{narayan2003,pu05}.  Roche-lobe overflow x-ray
binaries, however, might accrete a quite irregular field geometry.
The field geometry that arrives at the black hole, after travelling
from the source of material (molecular torus, star(s), etc.) to the
black hole horizon, likely depends sensitively on the reconnection
physics.

The reason why each system has some observed Lorentz factor has not
been well-understood.  One key idea of this paper is that the terminal
Lorentz factor is determined by the toroidal magnetic energy per unit
pair mass density energy near the location where pairs can escape to
infinity (beyond the so-called ``stagnation surface'').  Put another
way, the Lorentz factor is determined by the energy flux per unit rest
mass flux for the rest-mass flux in pairs beyond the stagnation
surface.  For GRBs, neutron diffusion is crucial to explain (and
limit) the Lorentz factor.  For AGN and x-ray binaries, since a
negligible number of baryons cross the field lines, pair-loading is
crucial to determine the Lorentz factor of the Poynting-dominated jet
since this determines the rest-mass flux or density.

\begin{figure}
\includegraphics[width=3.3in,clip]{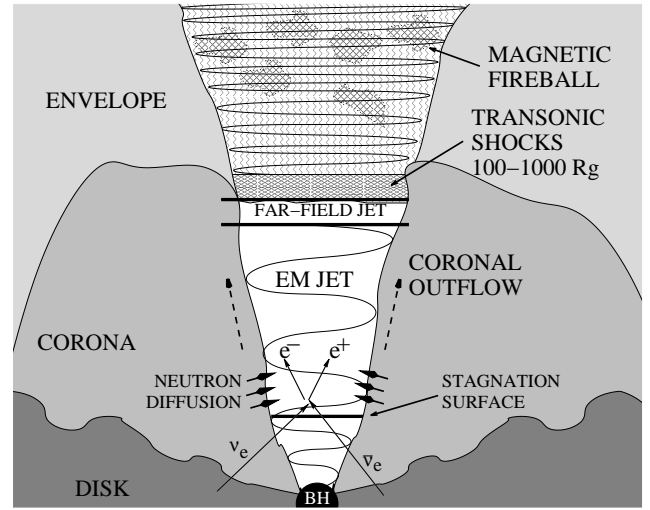}

\caption{Schematic of pair-production model and subsequent magnetic
fireball formation for GRB disks. Fireball is extremely optically
thick. Below a stagnation surface, pairs are accreted by the black
hole and so do not load the jet. Here $Rg=GM/c^2$. }
\label{model}
\end{figure}

\begin{figure}
\includegraphics[width=3.3in,clip]{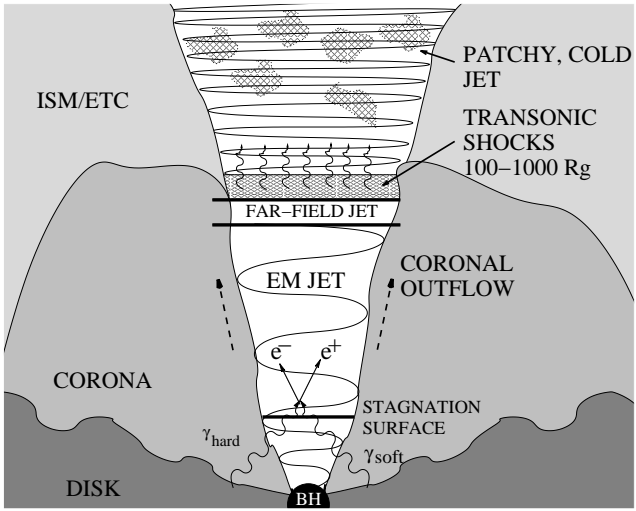}

\caption{Schematic of pair-production model and subsequent
shock-heating and emission. AGN jet is optically thin and emits
nonthermal and thermal synchrotron, while x-ray binary jet can be
marginally optically thick and emit via self-absorbed synchrotron and
by severe Compton drag.  Severe Compton drag can lead to destruction
of the Poynting-dominated jet. }
\label{modelagn}
\end{figure}

Figure~\ref{model} shows the basic picture for GRB systems, while
figure~\ref{modelagn} shows the basic picture for AGN and x-ray binary
systems.  An accreting, spinning black hole creates a magnetically
dominated funnel region around the polar axis.  The rotating black
hole drives a Poynting flux into the funnel region, where the Poynting
flux is associated with the coiling of poloidal magnetic field lines
into toroidal magnetic field lines.  The accretion disk emits
neutrinos in a GRB model ($\gamma$-ray and many soft photons for AGN
and x-ray binaries) that annihilate and pair-load the funnel region
within some ``injection region.''  For GRB systems, neutrons
Fick-diffuse across the field lines and collisionally decay into an
electron-proton plasma.

Many pairs (any type) are swallowed by the black hole, but some escape
if beyond some ``stagnation surface,'' where the time-averaged
poloidal velocity is zero and positive beyond.  Pairs beyond the
stagnation surface are then accelerated by the Poynting flux in a
self-consistently generated collimated outflow.  In the
electromagnetic (EM) jet, the acceleration process corresponds to a
gradual uncoiling of the magnetic field and a release of the stored
magnetic energy that originated from the spin energy of the black
hole.

One key result of this paper is that the release of magnetic energy
need not be gradual once the toroidal field dominates the poloidal
field, in which case pinch (and perhaps kink) instabilities can occur
and lead to a nonlinear coupling (e.g. a shock) that converts Poynting
flux into enthalpy flux \citep{e93,begelman1998}.  In the proposed GRB
model, this conversion reaches equipartition and the jet becomes a
``magnetic fireball,'' where the toroidal field instabilities drive
large variations in the jet Lorentz factor and jet luminosity.

In AGN systems, nonthermal synchrotron from shock-accelerated
electrons and some thermal synchrotron emission releases the shock
energy until the synchrotron cooling times are longer than the jet
propagation time.  For AGN, jet acceleration is negligible beyond the
{\it extended} shock zone, as suggested for blazars beyond the
``blazar zone'' \citep{sikora2005}. In x-ray binary systems, the shock
is not as hot and also unlike in the AGN (at least those like M87)
case the jet can be optically thick.  Thus these x-ray binary systems
self-absorbed synchrotron emit if they survive Compton drag.

For all these systems, at large radii patches of energy flux and
variations in the Lorentz factor develop due to toroidal
instabilities.  These patches in the jet could drive internal shocks
and at large radii they drive external shocks with the surrounding
medium.  The EM jet is also surrounded by a mildly relativistic matter
coronal outflow/jet/wind, which is a material extension of the corona
surrounding the disk.  This Poynting-baryon, coronal outflow
collimates the outer edge of the Poynting-dominated jet, which
otherwise internally collimates by hoop stresses.  The luminosity of
the Poynting-baryon jet is determined, like the Poynting-dominated
jet, by the mass accretion rate, disk thickness, and black hole spin.

This model is studied analytically in this paper, while in a companion
paper we study this model numerically using axisymmetric,
nonradiative, GRMHD simulations to study the self-consistent process
of jet formation from black hole accretion systems
\citep{mckinney2005b}.  Those simulations extend the work of
\citet{mg04} by including pair creation (and an effective neutron
diffusion for GRB-type systems) to self-consistently treat the
creation of jet matter, investigating a larger dynamic range in
radius, and presenting a more detailed analysis of the
Poynting-dominated jet structure.

Unless explicitly stated, the units in this paper have $G M = c = 1$,
which sets the scale of length ($r_g\equiv GM/c^2$) and time
($t_g\equiv GM/c^3$).  The mass scale is determined by setting the
(model-dependent) observed (or inferred for GRB-type systems) mass
accretion rate ($\dot{M}$[$g~s^{-1}$]) equal to the accretion rate
through the black hole horizon as measured in a simulation.  So the
mass is scaled by the mass accretion rate at the horizon, such that
$\rho_{0,disk}\equiv \dot{M}[r=r_H] t_g/r_g^3$ and the mass scale is
then just $m\equiv \rho_{0,disk} r_g^3 = \dot{M}[r=r_H] t_g$.  Unless
explicitly stated, the magnetic field strength is given in
Heaviside-Lorentz units, where the Gaussian unit value is obtained by
multiplying the Heaviside-Lorentz value by $\sqrt{4\pi}$.

The value of $\rho_{0,disk}$ can be determined for different systems.
For example, a collapsar model with $\dot{M}=0.1\msun s^{-1}$ and
$M\approx 3\msun$, then $\rho_{0,disk}\approx 3.4\times 10^{10}{\rm
g}\cm^{-3}$.  M87 has a mass accretion rate of $\dot{M}\sim
10^{-2}\msun\yr^{-1}$ and a black hole mass of $M\approx 3\times
10^9\msun$ \citep{ho99,reynolds96} giving $\rho_{0,disk}\sim 10^{-16}
{\rm g}\cm^{-3}$.  GRS 1915+105 has a mass accretion rate of
$\dot{M}\sim 7\times 10^{-7}\msun\yr^{-1}$ \citep{mr94,mr99,fb04} with
a mass of $M\sim 14\msun$ \citep{greiner2001a}, but see
\citet{kaiser04}.  This gives $\rho_{0,disk}\sim 3\times 10^{-4}{\rm
g}\cm^{-3}$.  This disk density scales many of the results of the
paper.

\section{Breakdown of ideal-MHD}\label{nonidealmhd}

Pair creation is critical to understand the physics of the highly
magnetized, evacuated funnel region that is associated with a
Poynting-dominated jet.  Pair creation is often invoked in order to
use the force-free electrodynamics or ideal MHD approximation in a
black hole magnetosphere (see, e.g., \citealt{bz77}).  However, in MHD
where rest-mass is treated explicitly, pair creation is not simply a
passive mechanism to short out spark gaps, which is the mechanism
invoked to allow the use of the force-free approximation.

Pair creation (and neutron Fick diffusion for GRB-type systems)
determines the matter flow in the magnetosphere, and thus the
matter-loading of any Poynting jet that emerges
\citep{phi83,punsly1991,lev05}.  As shown below, these sources of mass
loading self-consistently determine the Lorentz factor of the
Poynting-dominated jet and allows one to understand why black hole
accretion systems, while following the mass-invariant GRMHD equations
of motion, show a variety of jet Lorentz factors.

For GRBs, the radiative physics and neutron diffusion is shown to
determine the Lorentz factor of the Poynting-dominated jet.  For AGN
and x-ray binaries, the radiative physics is shown to determine the
Lorentz factor of the Poynting-dominated jet by determining its energy
and mass-loading.

The ideal MHD approximation (or force-free approximation in
magnetically dominated regions) has been shown to be a reasonably
valid theoretical framework to describe most of the nonradiative
dynamically important accretion physics around a black hole in GRBs,
AGN, and black hole x-ray binary systems \citep{phi83,mckinney2004}.
This approximation is the foundation of most studies of jets and
winds.  The ideal MHD approximation is a good approximation to
describe these flow properties except 1) in current sheets, which is
not treated explicitly in this paper ; 2) where pair creation
contributes a nonnegligible amount of rest-mass, internal energy, or
momentum density; 3) if the Goldreich-Julian (GJ) charge density is
larger than the number density of charge carriers ; and 4) if the
rest-mass flux due to ambipolar and Fick diffusion is negligible.

The first goal is to show that radiative annihilation into pairs
establishes a density of pairs much larger than the Goldreich-Julian
density.  The Goldreich-Julian charge density is never reached because
pair creation is completely dominated by neutrino annihilation in
GRB-type systems and photon annihilation in AGN and x-ray binary
systems.

Notice that the breakdown of ideal MHD is {\it required} in order to
extract black hole spin energy from a stationary, axisymmetric system.
\citet{wald74} showed that a rotating black hole induces a parallel
electric current in the surrounding magnetosphere such that the plasma
becomes nondegenerate (i.e. $E^i B_i \neq 0$).  \citet{bek78} argued
that if the ideal MHD approximation were valid, that no energy could
be extracted from a black hole.  This is because since $u^r < 0$ at
the horizon, and the radial energy flux can be written as $-T^r_t = E
\rho u^r$ (where $E$ is conserved along each flow line ; see
appendix~\ref{conservedflow}), then to extract net energy ($-T^r_t>0$)
from the black hole requires $E<0$.  However, in the ideal MHD
approximation $E>0$ at $r\sim \infty$, and by conservation of $E$
along each flow line, then $E>0$ on the horizon as well.  However,
based upon arguments by \citet{gj69}, \citet{bz77} argued that as the
magnetosphere is evacuated to the Goldreich-Julian charge density, the
parallel electric current separates the charges.  The Goldreich-Julian
rest-mass density for a species of electrons is
\begin{equation}\label{gjdensity}
\rho_{GJ} \sim m_e \frac{\Omega_H B}{2\pi c q} ,
\end{equation}
where $B$ is the magnetic field strength and $q$ is the electron
charge.  Once the parallel electric current is sufficiently large,
electrons are accelerated across the potential gap and photons can be
emitted by curvature radiation or inverse Compton scattering.  These
high energy photons either self-interact or are involved in a magnetic
bremsstrahlung interaction, ultimately leading to electron-positron
pairs.  These pairs would continuously short the induced potential
difference.  However, this picture does not establish how the
resulting pair plasma flow behaves.

Why have ideal GRMHD numerical models demonstrated the
Blandford-Znajek effect \citep{koide2002,mg04,dv05a,kom05} ?  These ideal
GRMHD numerical models implicitly break the ideal MHD approximation in
the required way to allow the extraction of energy from the black
hole.  For all the initial conditions and field geometries explored by
\citet{mg04} using the ``ideal'' GRMHD numerical model of an accreting
black hole, they always find that a highly magnetized polar region
forms and any material in this magnetosphere is either rapidly driven
into the black hole or driven out in a wind or jet.  They find that
strong field lines tie the black hole horizon to large radii.  Thus,
necessarily these ideal MHD models break the ideal MHD approximation
at a stagnation point where the poloidal velocity $u^p=0$.
Necessarily matter is created (at least) in this location since matter
inside this surface goes into the black hole and matter beyond it goes
away from the black hole.  This aspect is similar to the
charge-starved magnetosphere models where there is a spark-gap
\citep{rs75} where particles are generated (for a review see
\citealt{lev05}).  Once the magnetosphere reaches an axisymmetric,
quasi-stationary state, then the departure from the ideal-MHD
condition can be measured as deviations from conservation of the
conserved flow quantities given in equations~\ref{OMEGAF} to
equations~\ref{LCONS}.

Notice that for a realistic accretion disk the BZ power is different
than the typically used estimates \citep{mckinney2005a}.  For
$j\gtrsim 0.5$, they find that the efficiency in terms of the mass
accretion rate is
\begin{equation}\label{EMTOTEFF}
\eta_{EM,tot}=\frac{P_{tot}}{\dot{M}c^2} \approx 15\% \left(\frac{\Omega_H}{\Omega_H[j=1]}\right)^4 ,
\end{equation}
and
\begin{equation}\label{EMJETEFF}
\eta_{EM,jet}=\frac{P_{jet}}{\dot{M}c^2} \approx 7\% \left(\frac{\Omega_H}{\Omega_H[j=1]}\right)^5 ,
\end{equation}
where $r_g=GM/c^2$, $\Omega_H=jc/(2r_H)$ is the rotation frequency of
the hole, $r_H=r_g(1+\sqrt{1-j^2})$ is the radius of the horizon for
angular momentum $J=j GM^2/c$, and $j=a/M$ is the dimensionless Kerr
parameter, where $-1\le j\le 1$. However, net electromagnetic energy
is not extracted for $j\lesssim 0.5$ (including retrograde accretion)
when an accretion disk is present \citep{mg04}.  This high efficiency
is a result of the near equipartition of the magnetic field strength
($(B^r)^2$) in the polar region at the horizon and the rest-mass
density in the disk at the horizon.  If the black hole has $j\approx
0.9$, then $\approx 1\%$ of the accreted rest-mass energy is emitted
back as Poynting flux in the form of a jet and $\approx 3\%$ is
emitted back in total (so obviously $2\%$ goes into the disk and
corona -- about equally it turns out).

\subsection{GRB Pair Creation Model}\label{GRBPAIRS}

In GRB models, such as the collapsar model, neutrino/anti-neutrino
annihilation provides a source of electron-positron pairs at a much
larger density than the Goldreich-Julian density and so the
magnetosphere is not charge starved.  The cross-field magnetic
diffusion for charged species is negligible in such systems.  However,
free neutrons diffuse across the field lines and load the jet with an
electron-proton plasma \citep{le2003}, and this effect is considered
in the next section.

For the collapsar model, the jet has $B\sim 3\times 10^{15}{\rm
Gauss}$ \citep{mckinney2004} and $j\sim 0.9$, which gives
\begin{equation}\label{gjcollapsedensity}
\rho_{GJ}\sim 10^{-9} g cm^{-3}~~~{\rm(Collapsars)}.
\end{equation}

One can compare this to the density of pairs produced by neutrino
annihilation for the GRB collapsar model.  One can use the results in
table 3 and figure 9b of~\citet{pwf99} and the results in table 1 in
\citet{mw99}, which are fairly well fit to power laws, such that for
models with $\dot{M}\leq 0.1\msun s^{-1}$, and $M=3\msun$
\begin{eqnarray}\label{nueff}
\eta_{\nu\bar{\nu},ann} & \equiv &
\frac{L_{\nu\bar{\nu},ann}}{\dot{M}c^2} \nonumber\\
& \sim & 1\%
\left(\frac{\alpha}{0.1}\right)\left(\frac{j}{0.9}\right)^{7}\left(\frac{\dot{M}}{0.1\msun
s^{-1}}\right)^{3.8}
,
\end{eqnarray}
where this fit is based on an average between the conservative and
optimistic models of \citet{mw99}.  This assumes an average neutrino
energy of $\sim 10$MeV from the disk.  This says that for the collapsar
model with $j=0.9$ that about $1\%$ of the rest-mass accreted is given
back as positron-electron pairs due to neutrino annihilation, which is
similar to the Poynting flux from the black hole that goes into the
jet region as from equation~\ref{EMJETEFF}.

Published results of neutrino annihilation rates as a function of
position \citep{pwf99} can be used to obtain a preliminary model of
pair creation and incorporated into a GRMHD model.  The details of
this preliminary model end up not affecting the results, and a more
self-consistent model is left for future work.  The results primarily
depend on the overall annihilation luminosity and the basic radial
dependence of the energy injected as pairs.

Figure 9 and table 3 in \citet{pwf99} and table 1 of \citet{mw99} can
be used to obtain approximate radial and height dependent fits of the
energy density rate of depositing pairs into the jet region.  Their
figure 9 shows that the height and radial dependence of the pair
annihilation luminosity per unit distance.  These follow approximate
power laws or exponential laws for $j\gtrsim 0.2$.  A reasonable fit
is that
\begin{equation}
P[R]\approx e^{-\frac{R-r_g}{2.4 r_g}}
\end{equation}
and
\begin{equation}
Q[z]\approx e^{-\frac{z-r_g}{3.5 r_g}}
\end{equation}
for the luminosity per unit distance.  The coefficient is determined
by the total annihilation luminosity ($L_{\nu\bar{\nu},ann}$).  As in
\citet{pwf99}, photon null geodesic transport in curved spacetime is
neglected such that
\begin{equation}\label{lannint}
L_{\nu\bar{\nu},ann} \sim 2\pi A\int_{r_g}^\infty \int_{\theta=0}^{arctan[H/R]}P[R] Q[z]
r^2\sin{\theta} d\theta dr .
\end{equation}
Figure 6 of \citet{pwf99} can be used to obtain the disk thickness to
radius ratio
\begin{equation}
H/R\sim 0.1\left(\frac{r}{2r_g}\right)^{2/3}
\end{equation}
for $\dot{M}=0.1\msun s^{-1}$.  This allows one to determine that
$A\approx L_{\nu\bar{\nu},ann}/191$.  Thus the energy generation rate
can be written as
\begin{equation}\label{edotann}
\frac{\dot{e}_{\nu\bar{\nu},ann}}{\dot{\rho}_{0,disk}c^2} \sim
\frac{\eta_{\nu\bar{\nu}}}{N_A}
P[R] Q[z]~~~{\rm(Collapsars)} ,
\end{equation}
where $N_A\approx 191$ and we have defined $\dot{\rho}_{0,disk}\equiv
\dot{M}/r_g^3$.  However, the above $H/R$ assumes the jet fills around
the disk.  Rather, there is likely a thick corona between the disk and
jet \citep{mg04}.  Motivated by those simulations and the simulations
discussed in this paper, the jet region is presumed to exist within
\begin{equation}
\theta_j\approx 1.0\left(\frac{r}{3r_g}\right)^{-1/3}
\end{equation}
for $r\lesssim 100r_g$.  In that case $N_A\approx 70$.  Thus, a
significant fraction of the pairs are absorbed by the corona.
However, the corona could also {\it contribute} significantly to
neutrino production \citep{rs2005}.  Thus, $70\lesssim N_A\lesssim
191$ are reasonable limits.

Some fraction of the total energy deposited goes into pair rest-mass,
pair internal energy, radiation, and pair momentum. Here $f_\rho$
denotes the fraction turned into lab-frame mass, $f_h$ the fraction
turned into lab-frame internal energy and radiation, and $f_m$ the
fraction turned into momentum energy.  Thus $1=f_\rho+f_h+f_m$.  The
pair rest-mass density creation rate is defined as
\begin{equation}\label{neutrinodensityrate}
\frac{\dot{\rho}_{e^- e^+}}{\dot{\rho}_{0,disk}} =
f_\rho
\frac{\dot{e}_{\nu\bar{\nu},ann}}{\dot{\rho}_{0,disk}} ,
\end{equation}
where $\rho_{e^- e^+}=\rho_{0,e^- e^+} u^t$.

One can obtain a rough density measure in the injection region by
assuming the characteristic time scale for moving the pairs once they
have formed is the light crossing time at the stagnation surface
$t_{stag}\sim t_g (r_{stag}/r_g)$. Then $\rho_{0,e^- e^+}\sim
\dot{\rho}_{0,e^- e^+}t_{stag}$.  With $\rho_{0,disk}\equiv
\dot{\rho}_{0,disk} t_{stag}$,
\begin{equation}\label{neutrinodensity}
\frac{\rho_{e^- e^+}}{\rho_{0,disk}} \sim f_\rho \left(\frac{\dot{e}_{\nu\bar{\nu},ann}}{\dot{\rho}_{0,disk}c^2}\right) \left(\frac{r_{stag}}{r_g}\right) .
\end{equation}

Notice that many pairs fall into the black hole, so do not contribute
to the jet rest-mass or energy.  Only those pairs beyond the
stagnation surface survive the gravity of the black hole, such that
the total annihilation luminosity into the jet is
\begin{equation}
L_{esc} =2\pi \int_{r=r_{stag}}^\infty
\int_{\theta=0}^{\theta_j}\dot{e}_{\nu\bar{\nu},ann} r^2 \sin{\theta} dr d\theta ,
\end{equation}
which is a similar integral as performed before.  However, notice that
particles injected with $r<r_{stag}$, by definition, fall into the
black hole since they are inside the stagnation surface where $u^p<0$.
This is unlike the BZ-power, which in steady state is well-defined and
conserved through the stagnation surface \citep{lev05}.  For
$r_{stag}=r_g$, all the injected mass reaches infinity by definition.
For $r_{stag}=10r_g$, only $11\%$ of the mass injected reach infinity.
Because any mass injected lost to the black hole is of no consequence
to the acceleration at large distances, then the true efficiency of
pairs that load the jet is
\begin{equation}\label{etapairs}
\eta_{esc} = \frac{L_{esc}}{\dot{M}c^2}
\end{equation}
rather than $\eta_{\nu\bar{\nu},ann}$ for all $r_{stag}$.  One can
show that $\eta_{esc}=\eta_{\nu\bar{\nu},ann}$ for $r_{stag}=r_g$, but
is reduced to $\eta_{esc} \approx 0.11 \eta_{\nu\bar{\nu},ann}$ for
$r_{stag}=10r_g$ due to the loss of pairs into the black hole.

The pairs annihilate after formed by neutrino annihilation.
Equation~\ref{taupa} gives the pair annihilation rate.  For GRB
models, such as the collapsar model, the pair annihilation timescale
is $t_{pa}\sim 10^{-16}{\rm s} \ll GM/c^3 \sim 10^{-5} {\rm s}$ and
similarly all along the jet.  Thus, the pairs annihilate and form a
thermalized fireball.  A fraction $f_\rho+f_h$ of the energy injected
is turned into a electron-positron pair-radiation fireball.  The
typical angle of scattering neutrinos gives $f_m\sim f_\rho+f_h$
\citep{pwf99}.  Thus, for the fireball formation region $f_\rho+ f_h
\sim f_m\sim 0.5$ within factors of a few, and this is independent of
the energy of the neutrinos or the efficiency of annihilation.

All of the mass energy thermalizes into the fireball with a
temperature given by equation~\ref{u0tot}.  The formation fireball
pair mass plus pair internal energy density plus radiation internal
energy is
\begin{equation}\label{neutrinoenergydensity}
\frac{q_{0,tot}u^t u_t+p_{e^- e^+}+p_{\gamma}}{\rho_{0,disk}c^2}
\sim \left(\frac{\dot{e}_{\nu\bar{\nu},ann}}{\dot{\rho}_{0,disk}c^2}\right) \left(\frac{r_{stag}}{r_g}\right) ,
\end{equation}
where $q_{0,tot}\equiv (\rho_{0,e^- e^+}c^2+u_{0,e^-
e^+}+u_{0,\gamma})$.  This equation connects the energy injection
process in terms of the GRMHD equations of motion for a given
energy-at-infinity injection rate.  See also appendix~\ref{GRMHD}.

Using the discussion here and the equations in appendix~\ref{GRMHD},
one can show that for $\alpha=0.1$, $\dot{M}=0.1\msun s^{-1}$, and
$j=0.9$, the fireball temperature is $T\sim 10^{10}{\rm K}\sim 1{\rm
MeV}$ in the injection region.  Thus pair rest-mass energy is nearly
in equipartition with the pair internal energy and radiation.  In
particular, $f_\rho\approx f_h/8.5$.  Since $f_\rho+f_h\sim 0.5$, then
$f_\rho\approx 0.05$, $f_h\approx 0.45$, and $f_m\sim 0.5$.

Beyond the initial fireball formation, the Boltzmann factor gives that
once the temperature drops below $T\sim 6\times 10^9$K the number of
pairs decreases exponentially with temperature.  However, the fireball
is optically thin only at much larger radii of $r\sim 10^8-10^{10}
r_g$.  So until that radius, the radiation provides an inertial drag
on the remaining pair plasma and the gas is radiation dominated.

For example, for $j=0.9$, $\dot{M}=0.1\msun s^{-1}$, $r_{stag}=6r_g$,
then the initial fireball rest-mass is
\begin{equation}
\frac{\rho_{e^- e^+}}{\rho_{0,disk}}\sim 10^{-6}~~~(r_{stag}=6r_g) .
\end{equation}
If instead the model were an $\alpha=0.01$ model, then the efficiency
would be about $10$ times less and the density ratio would be about
$10$ times less at
\begin{equation}\label{collapsardensity}
\frac{\rho_{e^- e^+}}{\rho_{0,disk}}\sim 10^{-7}~~~(\alpha=0.01) .
\end{equation}
In any case this is roughly
\begin{equation}
\rho_{e^- e^+}\gtrsim 10^3 g cm^{-3} ,
\end{equation}
which is about 12 orders of magnitude larger than the GJ density given
in equation~\ref{gjcollapsedensity}, and so the black hole is far from
starved of charges.

\subsection{GRB Baryon Contamination}

Neutron diffusion across field lines leads to baryon contamination of
the (otherwise) electron-positron-radiation jet.  The neutrons
Fick-diffuse \citep{le2003} or diffusion due to ambipolar diffusion
(see appendix~\ref{GRMHD}) across the field lines and baryon-load the
jet.  The neutrons undergo a fast collisional avalanche into protons
and electrons that are then carried along with the electromagnetic and
Compton-thick outflow.  As shown in appendix~\ref{GRMHD}, the mass
injection rate of neutrons (and so proton+electrons) is
\begin{equation}\label{fickmdotused2}
\dot{M}_{inj,Fick} \sim 7\times 10^{-5} \dot{M}_{acc} 
\end{equation}
where the density of electron-proton plasma is
\begin{equation}\label{fickrhoused}
\rho_{p e^-} \sim 3\times 10^{-7} \left(\frac{r}{r_g}\right)^{-4/3} \rho_{0,disk} .
\end{equation}

Notice that this mass injection rate and density are comparable to the
injection-region rest-mass density in electron-positron pairs for
$\alpha=0.01$.  As mentioned above, father out in the jet the
electron-positron pairs annihilate and contribute only an additional
$\sim 10\%$ to the internal energy.  Thus the total internal energy is
sufficient to describe the gas without including pair annihilation and
the rest-mass in baryons is sufficient to describe the gas rest-mass
for models with $\alpha=0.01$ and for all models for $r\gtrsim 10r_g$
where the pair-density is exponentially smaller than the baryon
density in the jet.  Thus, the injection of ``pairs'' described in the
previous section can also approximately account for the Fick diffusion of
neutrons.  This fact is exploited to simulate the collapsar model in
\citet{mckinney2005b}.

\subsection{AGN and X-ray Binaries Pair Creation Model}

In AGN and x-ray binaries, the pairs are produced by scattering of
$\gtrsim 1{\rm MeV}$ $\gamma$-rays with other photons (for a review
see \citealt{phi83} and chpt. 6, 9, and 10 in \citealt{punsly2001}).
These $\gamma$-rays could be produced by, for example, Comptonization
of disk photons through a gas of relativistic electrons, non-thermal
particle acceleration in shocks (see, e.g., \citealt{nishikawa2003}),
or reconnection events.  For example, the (non-radiative) simulations
of \citet{mg04} show an extended corona that could be source of
Comptonization.  They also find an edge between the corona and funnel
that contains frequent shocks with sound Mach number $M\sim 100$.
Also, they found reconnection is quite vigorous in the plunging region
at $r\sim 3-6r_g$, leading in part to the hot coronal outflow.

These sites of Comptonization, shocks, and reconnection are likely
sources of the requisite $\gamma$-rays.  In place of a detailed model
of these processes, it is assumed that some fraction ($f_\gamma$) of
the true bolometric luminosity is in the form of these $\gamma$-rays
that do collide with softer photons in the funnel region.  For a
bolometric luminosity $L_{bol}\sim \eta_{eff} \dot{M} c^2$, then the
annihilation efficiency is
\begin{equation}
\eta_{\gamma\gamma,ann} \equiv f_\gamma \frac{L_{bol}}{\dot{M}c^2}
= f_\gamma \eta_{eff} ,
\end{equation}
where $\eta_{eff}$ is the total radiative efficiency, which could
include emission from both the disk and jet near the base.  Notice
that $\eta_{eff}$ depends on the black hole spin, among other things.
However, the value of $\eta_{eff}$ is obtained from (model-dependent)
values for the mass accretion rate and bolometric luminosity.

Extrapolating from gamma-ray observations of black hole x-ray binary
systems suggests that in either the quiescent or outburst phase,
$f_\gamma \sim 1\%$ of the true bolometric luminosity is in the form
of $>1$MeV photons (see, e.g., \citealt{lw05}).  These are likely
produced quite close to the black hole.  For an injection region with
a half opening angle for the jet of $\theta_j\sim 57^\circ$
\citep{mg04}, about $(2\theta_j)^2/(4\pi)\approx 1/3$ of these photons
enter the jet region.  Thus, it is assumed that a large fraction of
these $>1$MeV photons give up their energy into producing pairs in the
funnel region with some fraction of the energy going into rest-mass
($f_\rho$).  These pairs do not annihilate so form a collisionless
plasma (see appendix~\ref{GRMHD}).  See also \citet{punsly2001} for
why $f_\gamma\sim 1\%$ is reasonable, based on assuming the infall
rate is equal to the pair creation rate.  In general $f_\gamma$
depends on the state of the accretion flow, and a self-consistent
determination is left for future work.

As in the collapsar case this gives us a density rate or a typical
density.  In this case the stagnation surface is close to the black
hole since the emission is likely always optically thin, thus
$r_{stag}=3r_g$.  The author knows of no calculations that give a
radial dependence for the annihilation energy generation rate.  A
reasonable choice is the same radial dependence as for neutrino
annihilation, since while the efficiency of neutrino scattering is
much lower, the radial structure is determined by a similar
calculation.  The typical thick disk ADAF model assumes the electrons
and protons are weakly coupled, which gives a disk thickness mostly
independent of radius such that $H/R\sim 0.85$ and is weakly dependent
on $\alpha$ \citep{ny95}.  For the AGN and x-ray binaries, jets are
presumed to occur in the presence of a thick (ADAF-like) disk close to
the black hole.  While observations show that the disk at $r\gtrsim
6r_g$ undergoes state transitions, steady jets are only observed in
the low-hard state when the disk is likely geometrically thick.  For
this $H/R$, equation~\ref{lannint} gives $A=L_{\gamma\gamma,ann}/53$
or $N_A=53$.  Now equations~\ref{neutrinodensityrate} and
\ref{neutrinodensity} also apply for photon annihilation but with
$N_A=53$.

For AGN accretion disks, \citet{phi83} already showed that $\gamma$
rays from the accretion disk corona interact with x-rays to produce
electron-positron pairs in sufficient density above the
Goldreich-Julian density.  For example, M87 has a nuclear bolometric
luminosity of $L_{\gamma}\approx 2\times 10^{42}\ergps$ and a mass
accretion rate of $\dot{M}\sim 10^{-2}\msun\yr^{-1}$
\citep{ho99,reynolds96} giving $\eta_{eff}\sim 4\times 10^{-3}$.
Unlike neutrinos, $\gamma$-rays are efficient at creating pairs and
most of the $\gamma$-rays are at around $1$MeV so the fraction of the
energy put into rest-mass is $f_\rho\sim 1$.  If $f_\gamma\sim 1\%$,
then
\begin{equation}
\rho_{e^- e^+}\sim 10^{-23} g cm^{-3}~~~(f_\gamma\sim 1\%).
\end{equation}
The field strength in M87 is $B\sim 0.1~{\rm to}~ 50{\rm Gauss}$
\citep{mckinney2004}.  For $j=0.9$ this gives that
\begin{equation}
\rho_{GJ}\lesssim 10^{-32} g cm^{-3}~~~{\rm (M87)}.
\end{equation}
This is about 9 orders of magnitude lower than the pair creation
established density, so the black hole is not charge starved.  For
M87,
\begin{equation}
\frac{\rho_{e^- e^+}}{\rho_{0,disk}}\sim 10^{-7}~~~{\rm (M87)}.
\end{equation}

For black hole x-ray binaries a similar calculation is performed.  For
example, GRS 1915+105 has a mass accretion rate of $\dot{M}\sim
7\times 10^{-7}\msun\yr^{-1}$ and a bolometric luminosity $L\sim
10^{40}\ergps$ \citep{mr94,mr99,fb04} with a mass of $M\sim 14\msun$
\citep{greiner2001a}, but see \citet{kaiser04}.  This gives
$\eta_{eff}\approx 0.26$.  If $f_\gamma\sim 1\%$ and $f_\rho\sim 1$,
then
\begin{equation}
\rho_{e^- e^+}\sim 10^{-9} g cm^{-3}~~~(f_\gamma\sim 1\%).
\end{equation}
The field strength is $B\sim 10^6{\rm Gauss}$ if the disk is in the
thick ADAF-like state \citep{mckinney2004}.  For $j=0.9$ this gives
that
\begin{equation}
\rho_{GJ}\sim 10^{-19} g cm^{-3}~~~{\rm (GRS1915+105)}.
\end{equation}
This is about 9-10 orders of magnitude larger than the GJ density, so
the black hole is not charge starved.  For GRS 1915+105,
\begin{equation}
\frac{\rho_{e^-e^+}}{\rho_{0,disk}}\sim 10^{-5}~~~{\rm (GRS1915+105)} .
\end{equation}

Notice that GRS 1915+105, and many x-ray binaries, are more
radiatively efficient than most AGN.  This means x-ray binaries have
jets loaded with more pair-mass density per unit disk density.  This
will impact the presence and speed of any Poynting-lepton jet, as
described in the next section.

Finally, clearly temporal variations in the disk structure and mass
accretion rate directly affect the actual pair creation rate in the
jet region.  A self-consistent treatment of this (time-dependent)
radiative physics is left for future work.

\section{Jet Lorentz Factor}\label{lorentzfactor}

The Lorentz factor of the jet can be measured either as the current
time-dependent value, or, using information about the GRMHD system of
equations, one can estimate the Lorentz factor at large radii from
fluid quantities at small radii.  The Lorentz factor as measured by a
static observer at infinity is
\begin{equation}
\Gamma\equiv u^{\hat{t}}=u^t\sqrt{-g_{tt}}
\end{equation}
in Boyer-Lindquist coordinates, where no static observers exist inside
the ergosphere.  This is as opposed to $W\equiv u^t\sqrt{-1/g^{tt}}$,
which is the Lorentz factor as measured by the normal observer as used
by most numerical relativists.

For a GRMHD model, to determine the Lorentz factor at $r\sim\infty$,
notice that equations~\ref{ECONS} and~\ref{LCONS} specify that $E$ and
$L$ are conserved along each flow line.  For flows with a magnetic
field that is radially asymptotically small compared to near the black
hole, $B_\phi\rightarrow 0$ as $r\rightarrow\infty$. Also, for a
nonradiative fluid, the internal energy is lost to kinetic energy, and
so $h\rightarrow 1$ as $r\rightarrow \infty$.  Thus
$\Gamma_\infty=-u_t[r=\infty]$.  Now, since $\Phi$ and $\Omega_F$ are
conserved along a flow line, then trivially
\begin{eqnarray}
\Gamma_{\infty}         & = & E = -h u_t    + \Phi\Omega_F B_\phi \\
u^{\hat{\phi}}_{\infty} & = & L = h u_\phi + \Phi B_\phi ,
\end{eqnarray}
where $h=(\rho_0+\IEDEN+p)/\rho_0$ is the specific enthalpy, $\Phi$ is
the conserved magnetic flux per unit rest-mass flux, $\Omega_F$ is the
conserved field rotation frequency, and $B_\phi$ is the covariant
toroidal magnetic field.  $E$ and $L$ simply represent the conserved
energy and angular momentum flux per unit rest-mass flux.  Notice the
matter and electromagnetic pieces are separable.

Since $E$ is the hydrodynamic plus magnetic energy flux per unit
rest-mass flux, $B_\phi\rightarrow 0$ simply represents the conversion
of Poynting flux into rest-mass flux and $h\rightarrow 1$ represents
conversion of thermal energy into rest-mass flux.  These are what
accelerate the flow.  Alternatively stated for the magnetic term, from
equation~\ref{ACCEM1}, the fluid is accelerated by the magnetic
toroidal gradients associated with the Poynting outflow.

A rough estimate of the Lorentz factor at infinity $\Gamma_\infty$ can
be estimated by assuming all the enthalpy flux or all the Poynting
flux is lost to rest-mass flux that reach infinity.  Then from
equation~\ref{ECONS0}, one can break up the matter and electromagnetic
numerator and average the numerator and denominator separately to
obtain
\begin{equation}\label{GAMMA0}
\Gamma_\infty 
=
\Gamma^{(MA)}_\infty + \Gamma^{(EM)}_\infty
\sim
\left( \eta_{esc} + \eta_{EM} \right)
\left(\frac{\dot{M}_{disk}}{\dot{M}_{esc}}\right) ,
\end{equation}
where here ``esc'' refers to those pairs that escape the black hole
gravity.  In steady-state, only those pairs beyond the stagnation
surface escape to feed the jet.

Before estimating the value of the Lorentz factor for various systems,
the toroidal field is shown to be the source of the acceleration in
ideal MHD.  This also allows a probe of the structure of the jet,
which is not possible in the above averages.  In ideal-MHD, $E$ and
$L$ in equations~\ref{ECONS} and~\ref{LCONS} give the Lorentz factor
at infinity and the angular momentum per particle at infinity.

The enthalpy is just
\begin{equation}
h=(\rho+u+p)/\rho=1+\frac{4u}{3\rho}=1+\frac{4(1-f_\rho)}{3f_\rho}
\end{equation}
for a relativistic gas of electron-positron pairs that have $f_\rho$
of energy into rest-mass and the rest into thermal or momentum energy.
Now the magnetic term is
\begin{equation}
\Phi\Omega_F B_{\phi} = \frac{B^\phi B_\phi}{\rho_0
  u^t(v^\phi/\Omega_F-1)}.
\end{equation}
Notice that $E$ diverges for $v^\phi = \Omega_F$ (or
$u^r=u^\theta=0$), where the ideal MHD approximation breaks down.
Thus, $\Gamma_{\infty}$ is also determined by the non-ideal MHD
physics of particle creation in that region.

For an extended particle creation region, $\Gamma_{\infty}$ depends on
the mass, enthalpy, and momentum injected as a function of radius and
$\theta$ in the jet region \citep{lev05,punsly2001}.  For a narrow
($\delta r\ll r_g$), yet distributed, particle creation region, then
plausibly ideal MHD is completely reestablished at a slightly larger
radius where $v^\phi\ll \Omega_F\approx \Omega_H/2$ \citep{mg04}.  In
this case $\Phi\approx -\rho\Omega_F u^t/B^\phi$ (or
$u^r/B^r\approx-\Omega_F u^t/B^\phi$).  Thus, the magnetic piece is
\begin{equation}
\Phi\Omega_F B_{\phi}\approx \frac{B^\phi B_\phi}{\rho},
\end{equation}
where $\rho=\rho_0 u^t$ is the lab-frame density.  Written in
Boyer-Lindquist coordinates in an orthonormal basis, then the magnetic
piece is
\begin{equation}
\Gamma_\infty^{(EM)}\approx\frac{(B^{\hat{\phi}})^2}{\rho},
\end{equation}
where $B^{\hat{\phi}}=\sqrt{g_{\phi\phi}}B^\phi$.  Thus, in
Boyer-Lindquist coordinates
\begin{equation}\label{sigmaphi}
\Gamma_\infty \sim 1+\frac{4(1-f_\rho)}{3f_\rho} + \frac{(B^{\hat{\phi}})^2}{\rho c^2}
\end{equation}
for $r>r_H$. Thus the fluid energy at infinity is due to conversion of
thermal energy and toroidal field energy into kinetic energy.  The
actual value of $\Gamma_\infty$ depends on how narrow is the injection
region and the location of the stagnation surface.  The relevant
magnetic field is the toroidal magnetic field strength beyond the
injection region where ideal MHD is mostly reestablished.  Notice that
$\sigma=b^2/\rho_0$ is often used to parameterize the Lorentz factor
for a magnetically dominated flow, while perhaps
equation~\ref{sigmaphi} is more useful.

The polar field on the black hole horizon is approximately monopolar
even for rapidly rotating holes with $j\lesssim 0.95$ \citep{mg04}.
The monopole field solution can then be used, when properly
normalized, to give the functional dependence of $B^{\hat{\phi}}$ near
the black hole.  Then one may use equation~\ref{sigmaphi} to obtain
the approximate terminal Lorentz factor.  First, the black hole emits
a Poynting flux given by equation~33 in \citet{mg04} with the use of
their section~2.3.2 for the magnetic field.  In an asymptotic
expansion, which is good to factors of 2 for any $j$ even at $r=3r_g$,
then
\begin{equation}
B^{\hat{\phi}}\approx -C\frac{j}{8r/r_g}\sin{\theta}.
\end{equation}
Modifications at higher black hole spin are within factors of $1.5$ in
magnitude and there is a factor of $\lesssim 1.5$ enhancement near the
polar axis compared to lower spin.  In order to use this simple BZ
monopole, the jet region's ``C'' coefficient can be obtained from
GRMHD numerical models \citep{mg04}.  For a $H/R\sim 0.2-0.4$ disk
model, they find that the normalization constant ``$C$'' in the BZ
monopole solution is
\begin{equation}
C\approx 0.7\sqrt{\rho_{0,disk}c}
\end{equation}
(see equations 47-49 in \citealt{mg04}).  This clearly states that the
toroidal field in the polar region is nearly in equipartition with the
rest-mass density in the disk.  One can plug this into
equation~\ref{sigmaphi}, but this equation would have limited
usefulness since the injection region is broad and the density is
difficult to estimate.  However, notice that $\Gamma_\infty$ has
angular structure if the magnetic term dominates since then
$\Gamma_\infty\propto \sin^2{\theta}$.  This is important to jet
structure described in \citet{mckinney2005b}.  Otherwise the result
has the same qualitative features as equation~\ref{GAMMA1}.

\subsection{Lorentz Factors in Collapsar Systems}\label{grblorentz}

This section shows that, without invoking super-efficient neutrino
emission mechanisms, only the Blandford-Znajek driven process can
drive the flow to the necessary minimal Lorentz factor to avoid the
compactness problem and unequivocally generate a GRB. Some GRBs
require up to $\Gamma\sim 500$ \citep{ls01}, so any invoked mechanism
must be able to explain this.  Some observation/models suggest some
bursts have even $\Gamma\sim 1000$ \citep{sr03}.

Equation~\ref{GAMMA0} can be written as
\begin{equation}\label{GAMMACOLLAPSAR0}
\Gamma_\infty = \left( \eta_{esc} + \eta_{EM,jet}
\right)\left(\frac{\dot{M}}{\dot{M}_{esc}}\right) .
\end{equation}
Equation~\ref{etapairs} gives $\eta_{esc}$, which accounts for pair
capture by the black hole.  Here $\eta_{EM}$ is given in
equation~\ref{EMJETEFF} and as generally noted before,
$\eta_{esc}=\eta_{\nu\bar{\nu},ann}$ for $r_{stag}=r_g$, but is
reduced to $\eta_{esc} \approx 0.11 \eta_{\nu\bar{\nu},ann}$ for
$r_{stag}=10r_g$ due to the loss of pairs into the black hole.  Based
upon GRMHD numerical models studied in \citet{mckinney2005b}, a likely
value is $r_{stag}=5r_g$, for which $\eta_{esc} \approx 0.5
\eta_{\nu\bar{\nu},ann}$.

GRB-type systems are different than AGN and x-ray binary systems,
because neutron diffusion baryon-loads the jet.  The indirect
injection of protons and electrons is shown to dominate the rest-mass
in the jet because the electron-positron pairs annihilate to
negligible rest-mass.  The Fick diffusion mass injection rate is given
by equation~\ref{fickmdot} and is
\begin{equation}\label{fickmdotused}
\dot{M}_{inj,Fick} \sim 7\times 10^{-5} \dot{M}_{acc} 
\end{equation}
\citep{le2003}, and so $\dot{M}_{esc}=\dot{M}_{inj,Fick}$.

Plugging in the efficiencies for neutrino annihilation
(equation~\ref{nueff}) and Poynting-dominated jet efficiency
(equation~\ref{EMJETEFF}) into equation~\ref{GAMMACOLLAPSAR0}, one
finds for $\dot{M}\leq 0.1\msun s^{-1}$ for $j\gtrsim 0.1$ that
\begin{equation}\label{GAMMACOLLAPSAR}
\Gamma_\infty
\sim
140 \left(g_{\nu\bar{\nu},ann,esc} + g_{EM,jet}\right)
\end{equation}
where
\begin{equation}
g_{\nu\bar{\nu},ann,esc}\equiv
\left(\frac{\eta_{esc}}{\eta_{\nu\bar{\nu},ann}}\right)\left(\frac{\alpha}{0.1}\right)\left(\frac{j}{0.9}\right)^{7}\left(\frac{\dot{M}}{0.1\msun
  s^{-1}}\right)^{3.8}
\end{equation}
where for $\dot{M}\gtrsim 0.1\msun s^{-1}$, the power of $3.8$ sharply
levels out to $0$ \citep{dimat02}.  Also,
\begin{equation}
g_{EM,jet}\equiv 7\left(\frac{j}{1+\sqrt{1-j^2})}\right)^5 .
\end{equation}
These $g$'s are just the normalized the efficiencies.  It turns out
that for $j\approx 0.9$ that the efficiencies are similar for the
shown normalization of parameters.  Thus, one might expect that they
contribute equally to the energy of the jet.

If $r_{stag}=r_g$ and $\alpha=0.1$, then for the collapsar model with
$\dot{M}=0.1\msun s^{-1}$ and $j=0.9$, then neutrino annihilation and
BZ power are equal and typically give $\Gamma\sim 100-140$, which is
sufficient to avoid the compactness problem for typical bursts.

However, $r_{stag}\approx 5r_g$ is more likely.  The neutrino-driven
term then gives $\Gamma_\infty\sim 70$ and is dominated by the
BZ-driven term that stays at $\Gamma_\infty\sim 100$.  This is a
result of the loss of pairs into the black hole. Notice that the
results of \citep{pwf99,dimat02} and others do not include Kerr
geometry to trace null geodesics.  While the efficiency for
annihilation should increase due to gravitational focusing, more pairs
are also absorbed by the black hole.  Thus, it is unlikely that
gravitational focusing helps the neutrino-driven mechanism.

Also, $\alpha=0.1$ probably considerably overestimates the viscous
dissipation rate of a true MHD flow.  Despite typically $\alpha=0.1$
being used by many authors studying viscous models of disks,
\citet{sp01} showed that $\alpha=0.01$ is more representative of MHD
disks near the black hole.  This implies that the energy generation
rate in the disk and the generation rate of neutrinos is lower by
about an order of magnitude.  For normal neutrinos this gives
$\Gamma_\infty\sim 14$, insufficient to explain GRBs.

Another problem for the neutrino annihilation mechanism is that, with
the inclusion of optically thick neutrino transport, the efficiency of
neutrino emission is another few times lower for the collapsar model
\citep{dimat02}.  This gives $\Gamma_\infty\sim 5$, clearly a serious
problem for the neutrino annihilation model of GRBs.

The annihilation efficiency approximately scales with the average
neutrino energy \citep{pwf99}.  One must invoke super-efficient
neutrinos with an average neutrino energy of $\sim 210$MeV
\citep{rs2005} in order to obtain a neutrino annihilation power
comparable to the BZ power.  However, this is near the peak neutrino
energy estimated to come from a hot corona \citep{rs2005}, and the
corona is not expected to be the dominant source of neutrinos, so the
average neutrino energy should be smaller.

Notice that choosing $j=1$ only increases the neutrino efficiency, and
so $\Gamma_\infty$, by a factor of $2$.  Such a black hole spin is
only achievable when $H/R\lesssim 0.01$ \citep{gsm04}, which is not
representative of any GRB model \citep{kohri2005}.

So, without evoking super-efficient neutrino emission mechanisms and
unreasonably large neutrino annihilation efficiencies based upon only
optically thin emission, one must turn to the Poynting flux to drive
the jet. For $j=0.9$ one has $\Gamma_\infty\sim 100$, sufficient to
avoid the compactness problem is many GRBs and only invokes an
obtainable black hole spin.  Unlike the neutrino annihilation case,
the BZ efficiency has been computed self-consistently from GRMHD
numerical models of GRB-type disks \citep{mckinney2005b}.

Just as the neutrino-case has a proposed super-efficient mechanism,
there are super-efficient magnetic models that would increase the
terminal Lorentz factor.  Eventually after accreting a large amount of
magnetic flux, the magnetic pressure dominates the ram pressure of the
accretion flow and suspends the flow
\citep{igumenshchev2003,narayan2003}.  GRMHD numerical models are
suggested to not have simulated for long enough to see this effect.
In this case a larger amount of magnetic flux threads the black hole.
In this magnetically arrested disk (MAD) model, the field strength is
{\it comparable} to the rest-mass density in this case rather than
only a fraction of it.  Then, the BZ efficiency is $100\%$ for $j=1$
and an estimate of the jet efficiency is about $50\%$, which is $\sim
10$ times larger than previously.  Thus potentially $\Gamma\sim 1000$
is achievable with $j=0.9$.  Only super-efficient neutrinos with
average energy $2100$MeV can obtain such a jet power.

At the moment there is an insufficient study of the neutrino
annihilation rates for disks that have large optically thick regions,
so $\Gamma_\infty$ is not directly estimated for BH-NS or NS-NS
collisions.  However, at higher mass accretion rates the neutrino
emission efficiency levels off rather than increasing \citep{dimat02},
suggesting these systems should also be dominated by Poynting-flux
energy like for collapsars.

In summary, the neutrino-driven mechanism is probably dominated by the
BZ power.  Corrections due to optically thick neutrino-transport, a
realistic choice for $\alpha$ based upon MHD models, and the loss of
pairs into the black hole are the primary reducing factors compared to
previous expectations.  However, the BZ-effect generates sufficiently
energetic GRBs while only invoking an obtainable black hole spin.

Also, while a hydrodynamic jet mixes and can destroy jet structure, an
electromagnetic jet can internally evolve and could have a
distribution of Lorentz factors in the jet.  This is what is found in
\citet{mckinney2005b}.  Thus, while the electromagnetic jet may on
average have $\Gamma\sim 100$, the core of the jet may have
$\Gamma\sim 1000$.  Thus, without invoking super-efficient mechanisms,
only an electromagnetic BZ-driven jet can explain all
observed/inferred GRB Lorentz factors.

Based upon the density scaling from the simulations in
\citet{mckinney2005b} that are summarized in section~\ref{summaryfit},
and based upon appendix~\ref{compton}, the fireball is optically thick
to Compton scattering until $r\sim 10^{8}-10^{10} r_g$.  Simulations
discussed in \citet{mckinney2005b} show that Poynting flux is
continuously converted to heat in shocks and so all the energy flux
leads to acceleration. Thermal acceleration occurs until the fireball
is optically thin.  This acceleration reaches $\Gamma\sim 100-1000$
before the internal shocks generate the nonthermal emission.

\subsection{Lorentz Factors in AGN and X-ray Binary Systems}

This section shows that the {\it disk+jet} radiative physics is
crucial to determine the Lorenz factor of jets. It is shown that
radiatively inefficient AGN, such as M87, should have jets with
$2\lesssim \Gamma\lesssim 10$, while radiatively efficient systems,
such as GRS 1915+105, may have jets with $\Gamma\lesssim 2$, but they
may be Compton dragged to {\it nonrelativistic} velocities.

For AGN and x-ray binaries equation~\ref{GAMMA0} can be written as
\begin{equation}
\Gamma_\infty = \frac{1}{f_\rho}\left( \eta_{esc} + \eta_{EM}
\right)\left(\frac{\dot{M}c^2}{L_{esc}}\right) ,
\end{equation}
for a fraction $\dot{M}_{esc}=f_\rho L_{esc}$ of mass that escapes the
gravity of the hole.  Equation~\ref{etapairs} shows that the
mass-energy loading of the jet is reduced, compared to the total
injected mass-energy, due to loss of pairs into the black hole.
Since $\eta_{esc}\equiv L_{esc}/(\dot{M}c^2)$, then
\begin{equation}\label{GAMMA1}
\Gamma_\infty \sim \frac{1}{f_\rho} \left(1 + \frac{\eta_{EM}}{\eta_{esc}}\right)
\end{equation}
Here $\eta_{EM}$ is given in equation~\ref{EMJETEFF}.  As noted
before, $\eta_{esc}=\eta_{\gamma\gamma,ann}$ for $r_{stag}=r_g$, but
is reduced to $\eta_{esc} \approx 0.11 \eta_{\gamma\gamma,ann}$ for
$r_{stag}=10r_g$ due to the loss of pairs into the black hole.

For these systems $r_{stag}\approx 3r_g$, since the disk is likely
optically thin and emits harder radiation closer to the black hole.
This gives $\eta_{pairs}/\eta_{ann}\sim 0.83$ for the ADAF model of
$H/R$.  Here the fraction of energy in rest-mass is $f_\rho\sim 1$ due
to the efficiency of photon-photon annihilation for $>1$MeV photons
off the plentiful softer photons.  As discussed before, observations
suggest that the fraction of bolometric luminosity in $\gamma$-rays is
$f_\gamma\sim 0.01$.  Since the radiative efficiency should also
depend on black hole spin, it is assumed a typical system has $j\sim
0.9$ since that is a plausible equilibrium spin \citep{gsm04}.  A
detailed model of the radiative efficiency as a function of black hole
spin is left for future work.

Equation~\ref{GAMMA1} then gives that
\begin{equation}\label{sigmaagnxrb}
\Gamma_\infty \sim
1
+
\left(\eta_{eff}\right)^{-1} .
\end{equation}
For M87, $\eta_{eff}=0.004$ so $\Gamma_\infty=250$, while for
GRS1915+105 $\eta_{eff}=0.26$ so $\Gamma_\infty=5$.

As cited in the introduction, the observed apparent Lorentz factor of
AGN jets is $\Gamma_\infty\lesssim 30$, while inferred Lorentz factor
of some AGN jets is $\Gamma_\infty\lesssim 200$, in basic agreement
with the above estimates.  However, in particular for M87 this is
rather large.  Also, the GRS1915+105 estimate is a bit large.

The key difference between the collapsar event and AGN or x-ray
binaries is that in the collapsar case the photon luminosity
(including Compton upscattered by $\Gamma^2$) is negligible compared
to the jet luminosity, but see \citet{ghis00,lazzati2004}.  In M87 the
bolometric luminosity $L_{bol}$ is almost that of the jet luminosity
$P_{jet}$.  In GRS 1915+105 the bolometric luminosity is {\it greater}
than the jet luminosity.

The below discussion is a preliminary check on how radiative processes
affect the above results.  The results of simulations from
\citet{mckinney2005b} are invoked in order to obtain the density and
magnetic structure of the Poynting-lepton jet as summarized in
section~\ref{summaryfit}.  The simulations also show that at $r\sim
10^2 - 10^4 r_g$, any remaining Poynting flux is shock-converted into
enthalpy flux until they are in equipartition.  A self-consistent
simulation with synchrotron emission would then show the continuous
loss of Poynting flux until the synchrotron cooling timescale is
longer than the jet propagation timescale.  This still suggests the
jet magnetic field is finally in equipartition, but that much of the
energy is lost and so cannot acceleration the jet.

The {\it jet}, rather than just disk, radiative physics is necessary
in order to explain why $\Gamma_\infty=250$ is not achieved in M87 and
$\Gamma_\infty=5$ is not achieved in GRS 1915+105.  For M87, much of
the Poynting energy that leads to $\Gamma_\infty\sim 250$ is converted
to heat by shocks, which is then lost to nonthermal and some thermal
synchrotron emission.  Thus the Lorentz factor of the Poynting-lepton
jet achieved by $r\sim 10^2 - 10^3 r_g$ is likely the maximum
obtainable.  Numerical models of M87 discussed in
\citet{mckinney2005b} show that $2\lesssim \Gamma\lesssim 10$.  The
Poynting-lepton jet in GRS 1915+105 is likely destroyed by Compton
drag or at best $\Gamma_\infty\lesssim 2$.  Other black hole accretion
systems with different mass accretion rates and radiative efficiencies
have to be independently checked.  The below discussion of the
disk+jet radiative physics should be considered preliminary since a
full radiative transport is necessary to obtain a completely
self-consistent solution.

\subsubsection{Jet Destruction by Bulk Comptonization}

The previous section showed that jets from AGN and x-ray binaries
survive loading by pair production from $\gamma$-ray photons, but the
produced jet may not survive Compton drag (bulk Comptonization) by the
relatively soft photons emitted by the disk.

The simulation-based results are used to determine the optical depth
as given in equations~\ref{tauperp} and~\ref{taupar} to compute the
perpendicular and parallel optical depths to Comptonization.  For M87
$\tau_{\parallel}\sim 6\times 10^{-6}$ and $\tau_{\perp}\sim 6\times
10^{-6}$ at $r\sim 5r_g$ (stagnation surface where jet starts) and
$\tau_{\perp}\sim 4\times 10^{-6}$ at $r=120r_g$.  For GRS 1915+105,
$\tau_{\parallel}\gtrsim 11$ and $\tau_{\perp}\lesssim 4$ (stagnation
surface) and $\tau_{\perp}\lesssim 3$ at $r=120r_g$.  Thus M87 is not
Compton dragged by the disk photons, while GRS 1915+105 is likely
strongly Compton dragged by photons that originate near the base of
the jet and travel up through the jet or across the jet, or by
synchrotron self-Compton drag.

A Compton-dragged jet has a limited Lorentz factor that reaches an
equilibrium between decelerating and accelerating radiative processes.
Here is it assumed that most of the disk emission is at the base of
the jet.  Then the relevant scenario for GRS 1915+105 is the one where
all disk seed photons that enter the jet are scattered.  An isotropic
disk luminosity $L_{bol}$ shining on a conical jet with half-opening
angle $\theta_j$ dumps a luminosity of $L_{seed} \sim \theta_j^2
L_{bol}/4$ into the jet.  The photons effectively mass-load the jet
and an equilibrium Lorentz factor is reached, where
\begin{equation}\label{gammacomptondrag}
\Gamma_\infty \sim \left(\frac{P_{jet}}{2 L_{seed}}\right)^{1/3} \sim
\left(\frac{2\eta_{EM,jet}}{\eta_{eff}\theta_j^2}\right)^{1/3}
\end{equation}
for a cold beam of electrons (see, e.g., \citealt{brod04}). The
thermal Lorentz factor is comparable to the bulk Lorentz factor, so
thermal corrections are not significant.  For GRS 1915+105 this gives
a {\it nonrelativistic} velocity ($\Gamma\sim 1$) for the jet if most
of the emission enters the base.  Only if most of the emission enters
far ($r\gtrsim 10^2r_g$) from the base is up to $\Gamma\sim 2$
possible.  Thus is unlikely, so the Poynting-lepton jet that forms in
radiatively efficient systems, such as GRS 1915+105, are Compton
dragged to {\it nonrelativistic} velocities. Clearly the Lorentz
factor is sensitive to the disk thickness, the emission from the disk,
and the structure of the jet.  Thus these estimates should be treated
as preliminary.  A self-consistent radiative transfer calculation is
left for future work.

In summary, a radiatively efficient system loads the jet with more
pairs from the larger number of $\gamma$-rays.  This sets the maximum
possible Lorentz factor to be smaller than for radiatively inefficient
systems.  For systems with relatively high density jets, such as X-ray
binaries, the larger radiative efficiency also leads to an optically
thick jet that can be Compton dragged.

\subsubsection{Jet Destruction by Pair Annihilation}

Equation~\ref{taupa} gives the pair annihilation rate.  For AGN, such
as M87, the pair annihilation timescale is $t_{pa}\sim 10^{11}{\rm s}
\gg GM/c^3 \sim 10^{4} {\rm s}$ and for a jet propagation time
$t_{jet}\sim r/c$, a lower limit is $t_{pa}/t_{jet}\gtrsim 10^7$ all
along the jet.  Thus, most pairs do not annihilate.  See
appendix~\ref{GRMHD} on how this affects the fluid approximation.  See
also \citet{ghis92}.

For X-ray binary GRS1915+105, $t_{pa}\sim 2\times 10^{-4}{\rm s}
\gtrsim GM/c^3\sim 7\times 10^{-5}{\rm s}$ and $t_{pa}/t_{jet}\gtrsim
2$ all along the jet.  Thus, some nonnegligible fraction of the pairs
annihilate.  This also contributes to the destruction of the
Poynting-lepton jet in X-ray binary systems since much of this
radiative energy is lost at $r\gtrsim 150r_g$ where the jet is
optically thin along the jet and $r\gtrsim 300r_g$ where the jet is
optically thin perpendicular to the jet.

\subsubsection{Heat Loss by Synchrotron Emission}

For M87 it was estimated that $\Gamma_\infty\sim 250$, which is
inconsistent with observations.  However, much of the Poynting energy
is converted to internal energy in shocks induced by toroidal field
instabilities \citep{mckinney2005b}.  Thus, the synchrotron cooling
time might be sufficiently fast to release this internal energy that
would otherwise accelerate the flow through thermal acceleration.  For
numerical models described in \citet{mckinney2005b} that correspond to
M87, an equipartition ``magnetic fireball'' forms between $10^2r_g$
and $10^3 r_g$.  If this energy could be released, then the shocks
would again resume and all the Poynting and thermal energy would be
lost.  For an equipartition magnetic fireball half the energy is
thermal, so the jet internal energy is $u_{e^- e^+}\sim 125 \rho_{0,
e^- e^+} c^2$ and so the thermal Lorentz factor is $\Gamma_e\sim 125$.
The synchrotron cooling time in the lab frame is
\begin{equation}
t_{syn}\sim \frac{\Gamma_{bulk} \Gamma_e m_e c^2}{P_{syn}}\sim 6\pi
\frac{\Gamma_{bulk} m_e c^2}{\Gamma_e \sigma_T c B^2} .
\end{equation}
This gives that
\begin{equation}
t_{syn}\sim 10^4 s \left(\frac{r}{r_g}\right)^{1.4}~~~(r<390r_g{\rm ,M87})
\end{equation}
and
\begin{equation}
t_{syn}\sim 1 s \left(\frac{r}{r_g}\right)^{3}~~~(r>390r_g{\rm ,M87}).
\end{equation}
For a typical lab frame jet propagation time of $t_{jet}\sim r/c$,
\begin{equation}
\frac{t_{syn}}{t_{jet}}\sim \left(\frac{r}{r_g}\right)^{0.4}~~~(r<390r_g{\rm ,M87})
\end{equation}
and
\begin{equation}
\frac{t_{syn}}{t_{jet}}\sim 10^{-4}
  \left(\frac{r}{r_g}\right)^{2}~~~(r>390r_g{\rm ,M87}).
\end{equation}
Hence, one would {\it not} expect synchrotron cooling to take much of
the internal energy away.

However, the ``magnetic fireball'' forms by shock heating and
electrons are dramatically accelerated in such relativistic
collisionless shocks. The shocks generate a power law (nonthermal)
distribution of electrons, where much of the energy is carried by
high-energy electrons
\citep{begelman84,be87,ach01,fm03,kw05,fender2005}.  Typically the
distribution is $N(E)\propto E^{-2.22}$.  For a pair plasma the
maximum energy is limited by synchrotron losses \citep{ach01}, and the
resulting synchrotron emission has photon energies of $E\sim
25\Gamma_e {\rm MeV}$.  This gives $\gamma$-ray and up to possibly TeV
emission beamed along the jet, as in blazars.  For example, Mrk 421
shows 15 minute variability, which for a mass of $1.9\times 10^8\msun$
would suggest an emission size on the order of the horizon size
\citep{punch92,gaidos96}.  However, relativistic time effects with
$\Gamma_{bulk}\sim 10$ place these emissions at $r\sim 10^2r_g$, which
coincides with the shock-heated transfast region discussed in
\citep{mckinney2005b}.

Thus, a significant portion of the shock-heated internal energy should
be emitted by shock accelerated electrons and lost through the
optically thin jet.  In the shocks, inverse Compton also contributes
to emission of high-energy photons and the loss of internal energy.
Shock-induced population inversions may generate cyclotron masers at
shock sites and lead to large brightness temperatures \citep{beg05}.

Thus, it is expected that much of the jet is cold with
$\Gamma_{bulk}\sim 5-10$ left over from pre-shock magnetic
acceleration.  As described in the simulations of
\citet{mckinney2005b}, patches of slightly faster or slower bulk
$\Gamma$ are present by $r\sim 10^2r_g$.  In M87-based models, these
range from $2\lesssim \Gamma\lesssim 10$.

The synchrotron emission angular frequency is
\begin{equation}
\omega_c\sim \frac{3\Gamma_e^2 q B \sin{\alpha}}{2m_e c},
\end{equation}
where $\sin{\alpha}\sim 1$.  This gives a characteristic synchrotron
frequency of
\begin{equation}
\nu_c\sim 3\times 10^{12}\left(\frac{r}{r_g}\right)^{-0.7}{\rm
  Hz}~~~(r<390r_g{\rm ,M87})
\end{equation}
and
\begin{equation}
\nu_c\sim 3\times 10^{14}\left(\frac{r}{r_g}\right)^{-1.5}{\rm
  Hz}~~~(r<390r_g{\rm ,M87}).
\end{equation}
For $r\sim 10^2 r_g$ where the fireball begins to form, this gives
$\nu_c\sim 100$GHz (radio).  By $r\sim 10^3 r_g$, $\nu_c\sim 10$GHz
(radio).  The emission frequency depends on the mass accretion rate
(and so $\rho_{0,disk}$) for any particular AGN.  As discussed in
\citet{mckinney2005b}, $r\sim 10^2r_g$ is also where the flow goes
superfast (supersonic).  Thus this is consistent with the idea that
the radio-bright static knots at the base of the jet in, for example,
Cen A is due to shocks in a transfast (transonic) transition
\citep{hard05}.

In summary, the jet in M87 likely emits most of the internal energy,
generated in shocks in the transonic transition, as nonthermal
synchrotron with some thermal synchrotron, such that the jet beyond
$10^3-10^4r_g$ is relatively cold with $2\lesssim
\Gamma_\infty\lesssim 10$.

Notice that in x-ray binaries, for example GRS 1915+105, have a jet
with $u/\rho_{0,e^- e^+}c^2\sim 2.5$ and so thermal $\Gamma_e\sim
2.5$.  This gives that
\begin{equation}
t_{syn}\sim 10^{-7} {\rm s}  \left(\frac{r}{r_g}\right)^{1.4}~~~(r<390r_g{\rm ,GRS})
\end{equation}
and
\begin{equation}
t_{syn}\sim 10^{-11}{\rm s} \left(\frac{r}{r_g}\right)^{3}~~~(r>390r_g{\rm ,GRS}).
\end{equation}
For a typical lab frame jet propagation time of $t_{jet}\sim r/c$,
\begin{equation}
\frac{t_{syn}}{t_{jet}}\sim 10^{-3}  \left(\frac{r}{r_g}\right)^{0.4}~~~(r<390r_g{\rm ,GRS})
\end{equation}
\begin{equation}
\frac{t_{syn}}{t_{jet}}\sim 10^{-7}
\left(\frac{r}{r_g}\right)^{2}~~~(r>390r_g{\rm ,GRS}) ,
\end{equation}
where GRS denotes GRS1915+105. Thus thermal synchrotron is
sufficiently fast to cool the jet.  Since the jet is optically thick,
as estimated above, then synchrotron self-absorption will dominate the
emission process, which is what is observed \citep{foster96,fb04}.
The thermal synchrotron emission has
\begin{equation}
\nu_c\sim 2\times 10^{15} \left(\frac{r}{r_g}\right)^{-0.7}{\rm
 Hz}~~~(r<390r_g{\rm ,GRS})
\end{equation}
and
\begin{equation}
\nu_c\sim 3\times 10^{17} \left(\frac{r}{r_g}\right)^{-1.5}{\textrm
  Hz}~~~(r>390r_g{\rm ,GRS}).
\end{equation}
Near the base this gives $0.01$keV emission (EUV).  These soft
synchrotron photons will be Compton upscattered (synchrotron
self-Compton) by the $\Gamma\lesssim 5$ jet to x-rays and contribute
to the destruction of the Poynting-lepton jet.  Like in AGN,
nonthermal synchrotron likely takes away much of the shock-generated
internal energy and this may account for some unidentified EGRET
sources.

It is beyond the scope of the present study to establish whether
nonthermal synchrotron, synchrotron self-Compton, or external
Comptonization accounts for most of the high-energy luminosity.

\subsubsection{Other issues}

Another possible way of contaminating the Poynting-dominated jet is by
accreting a complicated field geometry and so baryon-loading the polar
region.  This turns the jet into a mixed lepton-baryon Poynting jet.
Core-collapse presents the black hole with a field geometry that has
an overall single poloidal sign.  Compared to GRBs, AGN and x-ray
binary black holes are more likely to accrete nontrivial field
geometries leading to baryon contamination of the jet.  Especially in
Roche-lobe formed disks in x-ray binaries, it is likely that the
accreted field geometry is quite tangled, so the likelihood of a
relativistic Poynting jet is further reduced.

No black hole x-ray binary has been observed to have an
ultrarelativistic jet (V4641 Sgr is still not confirmed, but see
\citealt{chaty03}), despite the GRMHD physics in such systems being
identical and the Lorentz factor is otherwise independent of the mass
of the compact object.  However, due to their relatively high
radiative efficiency compared to AGN, x-ray binaries produce more
$\gamma$-ray flux that increases the pair loading for a given magnetic
field strength near the black hole.  Also, the relatively high
radiative efficiency means any Poynting-lepton jet is severely Compton
dragged since the jet is optically thick.  However, it is possible
that there exists a large population of low radiative efficiency
galactic black hole accretion systems.  These radiatively inefficient
systems would produce a large amount of Poynting flux per unit
rest-mass flux which would be shock-converted by toroidal field
instabilities into nonthermal synchrotron emission and could appear as
``microblazars.''  However, thus-far observed x-ray binaries should
not be as intrinsically luminous per rest-mass accretion rate since
the Poynting flux per rest-mass flux available to shock-heating is two
orders of magnitude smaller than available for AGN.  This is due to
the relatively high pair-loading in typical x-ray binaries.
Low-luminosity x-ray binaries would behave more like blazars, and so
low luminosity x-ray binary microblazars may account for some of the
unidentified EGRET sources.

\section{Relativistic Poynting-Baryon Jets}\label{mixed}

This section discusses how mildly relativistic Poynting-baryon jets
can explain many jet observations.  The origin of these jets is the
inner-radial accretion disk.  The origin of the mass is unstable
convective outflows and magnetic buoyancy, and the mass fraction
released is typically a few percent of the mass accretion rate
\citep{mg04,dv05a}.  The ratio of Poynting to baryon flux depends
mostly on the spin of the black hole \citep{pc90a,pc90b,mg04}.  Since
the Poynting flux from a rapidly rotating black hole that is absorbed
by the corona is also a few percent, the Poynting-baryon jet is
heavily baryon-loaded.  The heavy baryon-loading limits
Poynting-baryon jets to only mildly relativistic velocities.  The most
relativistic, collimated, and least baryon-loaded portion of the
Poynting-baryon jet is at the magnetic wall bounded by the
Poynting-dominated jet.

\subsection{Matter Jets and Outflows in AGN}

Most AGN should have Poynting-baryon jets.  This Poynting-baryon jet
may often lead to erroneous conclusions about the nature of the jet in
AGN systems.

For example, \citet{junor99,biretta99,biretta2002} suggest that M87
slowly collimates from about $60^\circ$ near the black hole to
$10^\circ$ at large distances.  However, two of their assumptions are
likely too restrictive.  First, they assumed the jet is always
conical, which is apparent from figure 1 in \citet{junor99}.  If the
jet is not conical this can overestimate the opening angle close to
the core (i.e. perhaps $35^\circ$ is reasonable all the way into the
core).  Second, their beam size is relatively large so that factors of
$2$ error in the collimation angle are likely. Finally, and most
importantly for this paper, they assumed that there is only one jet
component.  This likely leads to a poor interpretation of the
observations.  If there is a highly collimated relativistic
Poynting-lepton jet surrounded by a weakly collimated Poynting-baryon
jet, then this would also fit their observations.

Alternatively, if the accretion disk in M87 is a very thin SS-type
disk with $H/R\sim 0.00048$ \citep{mckinney2004}, then their
conclusion that there is slow collimation is plausible.  However, thin
disks may be much less efficient at producing jets \citep{lop99,ga97}
and may not be able to produce collimated jets
\citep{okamoto1999,okamoto2000}.  A form of the idea that winds
collimate jets has also been proposed by \citet{tb05} and applied to
M87, but they consider a model where the wind slowly collimates the
jet in order to fit observations.  Here we suggest that the
observations have been misinterpreted due to the presence of two
components: a well-collimated relativistic cold Poynting-lepton jet
and a mildly relativistic coronal outflow.  We suggest the broader
emission component is due to the coronal outflow.

Notice that more recent maps of the M87 jet-formation region show no
``jet formation'' structure \citep{krich04}.  Thus, the structures
seen previously may be transient features, such as associated with
turbulent accretion disk producing a dynamic coronal outflow.

Measurements of the apparent jet speed in M87 reveal typically
$\Gamma\sim 1.8$ near the core while $\Gamma\sim 6$ at larger radii.
However, some core regions are associated with $\Gamma\sim 6$ that
rapidly fade \citep{biretta99}.  This is consistent with a
two-component outflow where the cold fast moving core of the jet is
only observed if it interacts with the surrounding medium (or stars),
the slower coronal outflow, or it undergoes internal shocks.

For relatively thin disks or slowly rotating black holes,
Poynting-baryon jets could appear as ``aborted jets'' \citep{ghis04}.

The classical AGN unification models \citep{up95} invoke a dominant
role for the molecular torus and broad-line emitting clouds, while the
broad coronal outflow may significantly contribute to modifications
and in understanding the origin of the clouds
\citep{elvis2000,ernmfp04}.

Other erroneous conclusions could be drawn regarding the jet
composition.  Entrainment, which could occur at large distances when
the ideal MHD approximation breaks down, causes difficulties in
isolating the ``proper'' jet component's composition.  Worse is the
fact that there should be two separate relativistic jet components,
making it difficult to draw clear conclusions regarding the
composition \citep{guil83,cf93,lb96,sm00}.  However, is has been
recently suggested that only electron-positron jets could explain FRII
sources \citep{kt04}.

It is also often assumed that if the jet is highly collimated that it
is also highly relativistic near the black hole, which would suggest
Comptonization of disk photons should produce clear spectral features
\citep{sm00}.  However, the jet may rather slowly accelerate and
quickly collimate, which is universally what GRMHD numerical models
find.

\subsection{Jets in X-ray Binaries}

The results of the previous section suggest that the term
``microquasar'' does not accurately reflect the jet formation process.
If radiatively efficient systems have no Poynting-lepton jet, then
what produces their jets?  Mildly relativistic jets from black hole
microquasars may be produced by the inner-radial disk rather directly
by the black hole. The above results suggest that GRS1915+105 may not
have a Poynting-lepton jet during its quiescent accretion phase in the
low-hard state.  All black hole accretion systems with a thick disk
have a mildly relativistic $1\lesssim \Gamma\lesssim 3$ coronal
outflow due to convective instabilities and magnetic buoyancy
\citep{mg04}.  This component is sufficiently relativistic to explain
the jets from black hole (and most neutron star) x-ray binary systems.
This mechanism only requires a thick disk and not necessarily a
spinning black hole, where other unification models suggest that the
black hole spin is necessary \citep{meier2001}.

It has been suggested that the transient, more relativistic, jet
produced in GRS 1915+105 is the result the formation of a thin disk as
the ADAF collapses as the mass accretion rate increases \citep{fbg04}.
No particular model of the transient jet has been suggested.

Here we give a proposal for the disk-jet coupling in black hole x-ray
binary systems, such as GRS 1915+105.  In the prolonged hard x-ray
state the disk is ADAF-like and the system produces a Poynting
synchrotron self-absorbed jet with $\Gamma\lesssim 2$, which may be
partially or completely Compton dragged to nonrelativistic speeds.
However, in the thick state, a Poynting-baryon jet is produced with
$\Gamma\sim 1.5$.  During the soft state, the disk is SS-like
\citep{ss73} and the black hole polar field is relatively weak and the
system generates an uncollimated (more radial) weak optically thin
Poynting outflow.  During this phase there is also a weak
nonrelativistic uncollimated Poynting-baryon outflow.

During the transition between hard and soft x-ray states the
production of pairs decreases significantly in the funnel, but the
black hole polar magnetic field has yet to decay.  During this
transition, an optically thin Poynting-lepton jet with $\Gamma\sim
2-3$ is produced that is collimated by the remaining inner-radial
ADAF-like structure or the Poynting-baryon wind that was produced
prior.  The Lorentz factor produced in the transition depends on the
details of the disk structure, and so $\gamma$-ray emission, during
the transition.  Once the black hole field has decayed, the fast
transient jet shuts down.

Alternatively, during the transition to the high-soft state a
transient jet can emerge as the corona is suddenly exposed to more
Poynting flux from the black hole.  This last bit of coronal material
can be launched off as a faster transient baryon-loaded jet.  The
dynamics of the state transition is left for future work.  This
overall picture is in basic agreement with \citet{fb04}, with the
additional physics of pair creation dominating the Poynting-lepton jet
formation process.

It is interesting that the results of \citet{gd04} suggest that for at
least some black hole x-ray binaries that have jets, the black hole is
likely not rapidly rotating (i.e. perhaps $j\lesssim 0.5$).  For such
black holes, there is negligible Poynting flux in the form of a
Poynting-dominated jet \citep{mg04}.  Thus, our conclusion that black
hole x-ray binary jets are driven by coronal outflows is consistent
with the results of \citet{gd04}.  However, even if black hole x-ray
binaries were rapidly rotating they might not produce
Poynting-dominated jets.

SS443 is plausibly an $M\sim 20\msun$ black hole system that has a jet
with $v\sim 0.3c$ \citep{lopez2003}.  Such a low jet velocity can be
explained by a Poynting-baryon jet.  To explain the opening angle of
$\sim 1^\circ$ the disk should be very thick near the black hole,
while the pulsed jet features can be explained as an instability due
to the overly thick disk self-interacting at the poles near the black
hole.

\section{Summary of Companion Paper Numerical Results}\label{sumnumerical}

This section summarizes the results of \citet{mckinney2005b} using a
GRMHD code HARM \citep{gmt03} with an advanced inversion method
\citep{noble05}.  Kerr-Schild coordinates were used in order to avoid
numerical artifacts associated with causal interactions between the
inner-radial boundary and the rest of the flow.  Viscous models have
found this issue to be critical to avoid spurious fluctuations in the
jet \citep{mg02}, such as might be associated with codes using
Boyer-Lindquist coordinates.

\subsection{Jet propagation}

\begin{figure}
\subfigure{\includegraphics[width=3.3in,clip]{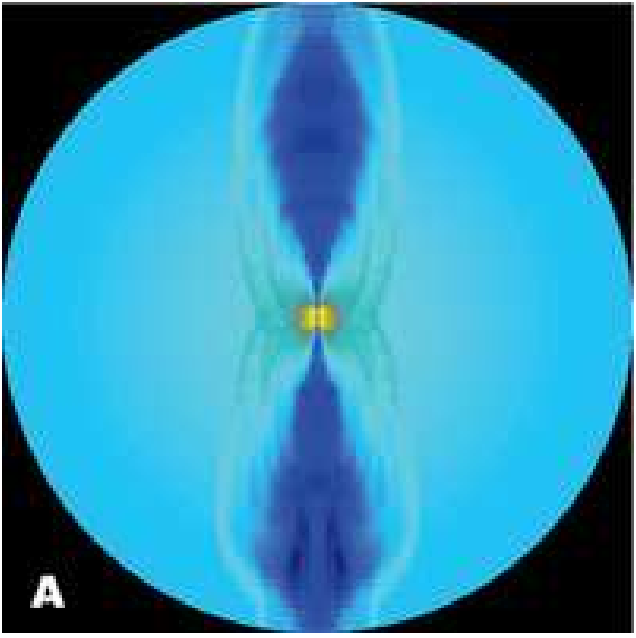}}
\subfigure{\includegraphics[width=3.3in,clip]{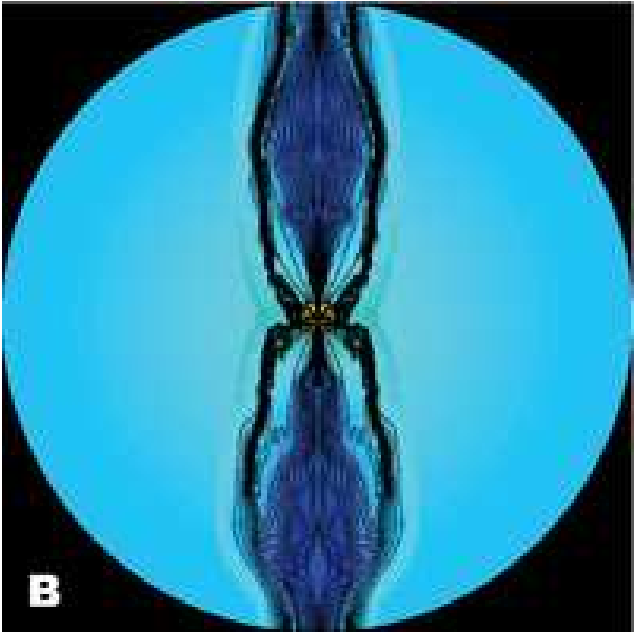}}

\caption{Jet has pummelled its way through presupernova core and
through $1/10$th of entire star.  Panel (A) shows final distribution
of $\log{\rho_0}$ on the Cartesian plane. Black hole is located at
center. Red is highest density and black is lowest. Panel (B) shows
magnetic field overlayed on top of log of density. Outer scale is
$r=10^4 GM/c^2$.}
\label{density}
\end{figure}

\begin{figure}
\subfigure{\includegraphics[width=3.3in,clip]{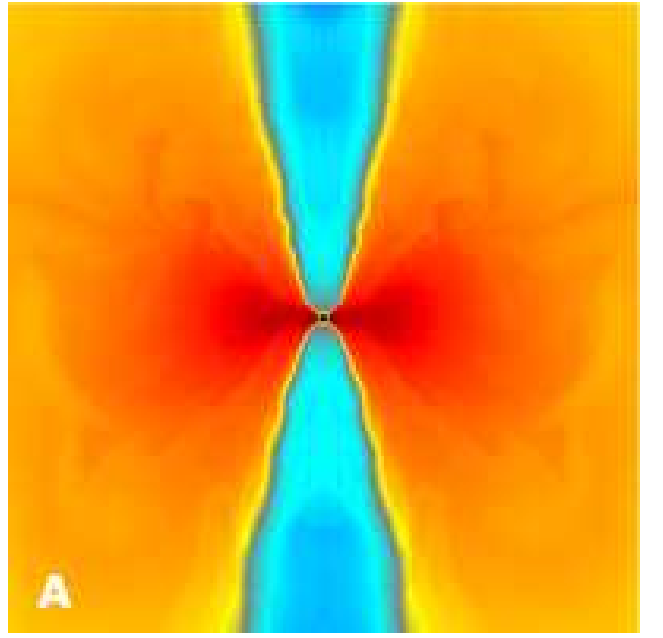}}
\subfigure{\includegraphics[width=3.3in,clip]{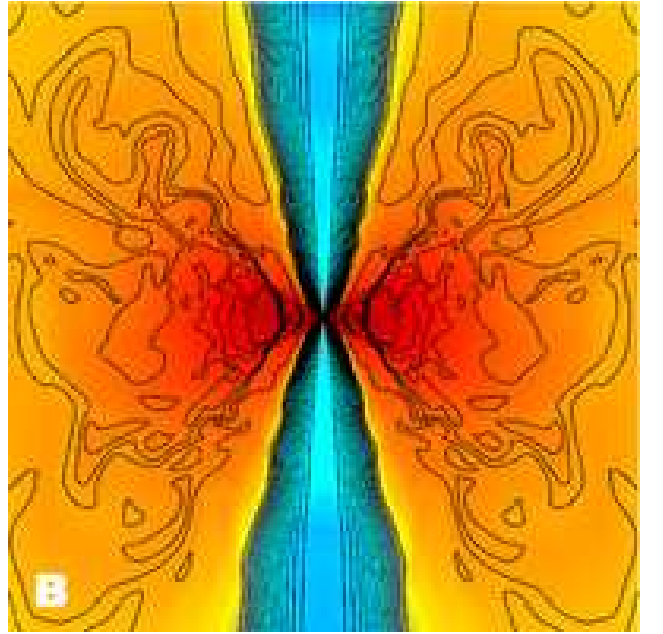}}

\caption{Strongest magnetic field near black hole in X-configuration
due to Blandford-Znajek effect and collimation of disk+coronal
outflow.  As in figure~\ref{density}, but outer scale is $r=10^2
GM/c^2$.  Black hole is black circle at center. Color scale is same as
in figure~\ref{density}. }
\label{densityzoom}
\end{figure}

As described in detail in \citet{mckinney2005b} and as shown in
figure~\ref{density} and~\ref{densityzoom}, the Poynting-dominated jet
forms as the differential rotation of the disk and the frame-dragging
of the black hole induce a significant toroidal field that launches
material away from the black hole by the same force described in
equation~\ref{ACCEM1}.

A coronal outflow is also generated between the disk and
Poynting-dominated jet.  In this model the coronal outflow has
$\Gamma_\infty\sim 1.5$.  The coronal-funnel boundary contains shocks
with a sonic Mach number of $M_s\sim 100$.  The inner-radial interface
between the disk and corona is a site of vigorous reconnection due to
the magnetic buoyancy and convective instabilities present there.
These two parts of the corona are about $100$ times hotter than the
bulk of the disk.  Thus these coronal components are a likely sites
for Comptonization and nonthermal particle acceleration.

Figure~\ref{density} and figure~\ref{densityzoom} show the final log
of density and magnetic field projected on the Cartesian z vs. x
plane.  For the purposes of properly visualizing the accretion flow
and jet, we follow \citet{mw99} and show both the negative and
positive $x$-region by duplicating the axisymmetric result across the
vertical axis.  Color represents $\log(\rho_0/\rho_{0,disk})$ with
dark red highest and dark blue lowest.  The final state has a density
maximum of $\rho_0\approx 2\rho_{0,disk}$ and a minimum of $\rho_0\sim
10^{-13}\rho_{0,disk}$ at large radii.  Grid zones are not smoothed to
show grid structure.  Outer radial zones are large, but outer $\theta$
zones are below the resolution of the figure.

Clearly the jet has pummelled its way through the surrounding medium,
which corresponds to the stellar envelope in the collapsar model. By
the end of the simulation, the field has been self-consistently
launched in to the funnel region and has a regular geometry there. In
the disk and at the surface of the disk the field is curved on the
scale of the disk scale height.  Within $r\lesssim 10^2r_g$ the funnel
field is ordered and stable due to the poloidal field dominance.
However, beyond $r\sim 10^2r_g$ the poloidal field is relatively weak
compared to the toroidal field and the field lines bend and oscillate
erratically due to pinch instabilities.  The radial scale of the
oscillations is $10^2 r_g$ (but up to $10^3r_g$ and as small as
$10r_g$), where $r\sim 10r_g$ is the radius where poloidal and
toroidal field strengths are equal.  By the end of the simulation, the
jet has only fully evolved to a state independent of the initial
conditions at $r\approx 5\times 10^3 r_g$, beyond which the jet
features are a result of the tail-end of the initial launch of the
field.  The head of the jet has passed beyond the outer boundary of
$r=10^4 r_g$.  Notice that the magnetic field near the black hole is
in an X-configuration.  This is due to the BZ-effect having power
$P_{jet}\propto \sin^2{\theta}$, which vanishes at the polar axis.
The X-configuration is also related to the fact that the disk+corona
is collimating the Poynting-dominated jet.  The field is mostly monopolar
near the black hole, and such field geometries {\it decollimate} for
rapidly rotating black holes in force-free electrodynamics
\citep{kras05}.

\subsection{Summary of Fits}\label{summaryfit}

A summary of the fits along a fiducial field line is given.    Near
the black hole the half-opening angle of the full Poynting-dominated jet
is $\theta_j\sim 1.0$, while by $r\sim 120r_g$, $\theta_j\sim 0.1$.
This can be roughly fit by
\begin{equation}
\theta_j\sim \left(\frac{r}{r_g}\right)^{-0.4}~~~{\rm (inner)}
\end{equation}
for $r<120r_g$ and $\theta_j\sim 0.14$ beyond.  The core of the jet
follows a slightly stronger collimation with
\begin{equation}
\theta_j\propto r^{-2/5}
\end{equation}
up to $r<120r_g$ and $\theta_j\sim 0.09$ beyond.  Also, roughly for
M87 and the collapsar model, the core of the jet has
\begin{equation}
\Gamma_{bulk}\sim \left(\frac{r}{5r_g}\right)^{0.44}~~~{\rm (inner)}
\end{equation}
for $5<r\lesssim 10^3r_g$ and constant beyond for the M87 model if
including synchrotron radiation, while the collapsar model should
continue accelerating and the power law will truncate when most of the
internal and Poynting energy is lost to kinetic energy and the jet
becomes optically thin at about $r\sim 10^9r_g$ or internal shocks
take the energy away.  If the acceleration is purely thermal without
any magnetic effect, then $\Gamma\propto r$ \citep{mr97}.  However, it
is not clear how the equipartition magnetic field affects the
acceleration.  Roughly for GRS 1915+105 the core of the jet has
\begin{equation}
\Gamma_{bulk}\sim \left(\frac{r}{5r_g}\right)^{0.14}~~~{\rm (inner)}
\end{equation}
for $5<r\sim 10^3r_g$ and constant beyond, with no account for Compton
drag or pair annihilation.  Also, for any jet system the base of the
jet has $\rho_0\propto r^{-0.9}~~~{\rm (inner)}$ for $r\lesssim 120
r_g$ and $\rho_0\propto r^{-2.2}~~~{\rm (outer)}$ beyond.  For the
collapsar and M87 models
\begin{equation}
\frac{\rho_0}{\rho_{0,disk}}\sim 1.5\times 10^{-9}
\left(\frac{r}{120r_g}\right)^{-0.9}~~~{\rm (inner)}
\end{equation}
and
\begin{equation}
\frac{\rho_0}{\rho_{0,disk}}\sim 1.5\times 10^{-9}
\left(\frac{r}{120r_g}\right)^{-2.2}~~~{\rm (outer)} ,
\end{equation}
while for GRS 1915+105 the inner-radial coefficient is $10^{-5}$ and
outer is $6\times 10^{-3}$. For the collapsar model, the inner radial
internal energy density is moderately fit by
\begin{equation}
\frac{u}{\rho_{0,disk}c^2}=4.5\times 10^{-9}\left(\frac{r}{120r_g}\right)^{-1.8}~~~{\rm (inner)} .
\end{equation}
The outer radial internal energy density is moderately fit by
\begin{equation}
\frac{u}{\rho_{0,disk}c^2}=4.5\times 10^{-9}\left(\frac{r}{120r_g}\right)^{-1.3}~~~{\rm (outer)}.
\end{equation}
The transition radius is $r\approx 120r_g$.  For M87 the internal
energy is near the rest-mass density times $c^2$ until $r\sim 120r_g$
when the dependence is as for the collapsar case.  For GRS1915+105 the
internal energy is near the rest-mass density times $c^2$ until $r\sim
120r_g$ and then rises to about $2.5$ times the rest-mass density
times $c^2$.  The inner radial toroidal lab field is well fit by
\begin{equation}
\frac{B^{\hat{\phi}}}{\sqrt{\rho_{0,disk}c^2}}[{\rm
    Gauss}]=0.0023\left(\frac{r}{390r_g}\right)^{-0.7}~~~{\rm (inner)}
\end{equation}
for $5<r<390r_g$. The outer radial toroidal lab field is well fit by
\begin{equation}
\frac{B^{\hat{\phi}}}{\sqrt{\rho_{0,disk}c^2}}[{\rm
  Gauss}]=0.0023\left(\frac{r}{390r_g}\right)^{-1.5}~~~{\rm (outer)}
\end{equation}
for $r>390r_g$.

For the typical jet with no atypical pinch instabilities, the energy
and velocity structure of the jet follow
\begin{equation}
\epsilon(\theta)= \epsilon_0 e^{-\theta^2/2\theta_0^2} ,
\end{equation}
where $\epsilon_0\approx 0.18$ and $\theta_0\approx 8^\circ$.  The
total luminosity per pole is $L_j\approx 0.023\dot{M}_0c^2$, where
$10\%$ of that is in the ``core'' peak Lorentz factor region of the
jet within a half-opening angle of $5^\circ$.  Also, $\Gamma_\infty$
is approximately Gaussian
\begin{equation}
\Gamma_\infty(\theta)= \Gamma_{\infty,0} e^{-\theta^2/2\theta_0^2} ,
\end{equation}
where $\Gamma_{\infty,0}\approx 3\times 10^3$ and $\theta_0\approx
4.3^\circ$.  Also, $\Gamma$ is approximately Gaussian
\begin{equation}
\Gamma(\theta)= \Gamma_0 e^{-\theta^2/2\theta_0^2} ,
\end{equation}
where $\Gamma_0\approx 5$ and $\theta_0\approx 11^\circ$.  The outer
sheath's ($\theta\approx 0.2$) seed photon temperature as a function
of radius is
\begin{equation}
T_{\gamma,seed}\sim 50{\rm keV} \left(\frac{r}{5\times
  10^3r_g}\right)^{-1/3} .
\end{equation}

\section{Discussion}\label{discussion}

For GRB jets, the injected Poynting flux probably dominates the
injected energy flux provided by neutrino annihilation.  This poses
problems for the classic neutrino-driven fireball model.
Super-efficient neutrino emission mechanisms with an {\it average}
neutrino energy of $210$MeV are required in order for the neutrino
annihilation energy to be as large as the energy provided by the BZ
effect.  However, the BZ effect itself might operate in a
super-efficient mode once flux has accumulated near the black hole
\citep{narayan2003}.  This vertical field threading the disk leads up
to $5$ times larger luminosity \citep{mg04}, in which case an average
neutrino energy of $1000$MeV is required to compete with the
BZ-effect.

For GRBs, equation~\ref{GAMMACOLLAPSAR} shows that slightly less
rapidly rotating black holes would generate slightly less Lorentz
factors but achieve a lower luminosity.  This is consistent with the
observation that harder long-duration bursts have higher luminosity,
and so suggests that small changes in the stellar rotation rate might
determine the hardness of long-duration bursts.

The fact that blazars are less luminous for increasing hardness could
be explained by the dependence on the jet Lorentz factor on the pair
creation physics.  Blazars could have similar black hole spin, but the
hardness of their emission is determined by the jet Lorentz factor.
Lower luminosity systems load the jet with less pairs and so the
Lorentz factor is larger.  Compton drag of environment or disk
reflected seed photons can then upscatter to very large energy, such
as observed in TeV-emitting BL-Lac objects.

Our results suggest that radiatively efficient x-ray binaries, such as
GRS1915+105, may only exhibit a relativistic Poynting-baryon jet.  In
particular, such a jet is relativistic only in the low-hard state when
the disk is geometrically thick.

\section{Conclusions}\label{conclusions}

Primarily two types of relativistic jets form in black hole (and
perhaps neutron star) systems.  The Poynting-dominated jet region is
composed of field lines that connect the rotating black hole to large
distances.  Since the ideal MHD approximation holds very well, the
only matter that can cross the field lines are neutral particles, such
as neutrinos, photons, and free neutrons.

The primary differences between GRBs, AGN, and black hole x-ray
binaries is the pair-loading of the Poynting-dominated jet, a similar
mass-loading by free neutrons in GRB-type systems, the optical depth
of the jet, and the synchrotron cooling timescale of the jet.

For GRB-type systems the neutron diffusion flux is sufficiently large
to be dynamically important, but small enough to allow $\Gamma\sim 100
- 1000$.  Beyond $r\sim 10r_g$ many of the electron-positron pairs
annihilate, so the Poynting-dominated jet is dominated in mass by
electron-proton pairs from collision-induced neutron decay.  Most of
the energy is provided by the BZ effect instead of
neutrino-annihilation.

For AGN and x-ray binaries, the density of electron-positron pairs
established near the black hole primarily determines the Lorentz
factor at large distances.  Radiatively inefficient AGN, such as M87,
achieve $2\lesssim \Gamma_\infty\lesssim 10$ and are synchrotron
cooling limited.  The lower the $\gamma$-ray radiative efficiency of
the disk, the more energy per particle is available in the shock-zone.
Radiatively efficient systems such as GRS1915+105 likely have no
Poynting-lepton jet due to strong pair-loading and destruction by
Comptonization by the plentiful soft photons for x-ray binaries with
optically thick jets.  However, all these systems have a mildly
relativistic, baryon-loaded jet when in the hard-low state when the
disk is geometrically thick, which can explain jets in most x-ray
binary systems.

In an companion paper \citet{mckinney2005b}, a GRMHD code, HARM, with
pair creation physics was used to evolve many black hole accretion
disk models.  The basic theoretical predictions made in this paper
that determine the Lorentz factor of the jet were numerically
confirmed.  However, Poynting flux is not necessarily directly
converted into kinetic energy, but rather Poynting flux is first
converted into enthalpy flux into a ``magnetic fireball'' due to shock
heating.  Thus, at large distances the acceleration is primarily
thermal, but most of that thermal energy is provided by
shock-conversion of magnetic energy.  In GRB systems this magnetic
fireball leads to thermal acceleration over an extended radial range.
The jets in AGN and x-ray binaries release this energy as synchrotron
and inverse Compton emission and so the jet undergoes negligible
thermal acceleration beyond $r\sim 10^2-10^3r_g$.

Based upon prior numerical \citep{mckinney2005b} and this theoretical
work, basic conclusions for collapsars include:
\begin{enumerate}
  \item Black hole energy, not neutrino energy, typically powers GRBs.
  \item Poynting-dominated jets are mostly loaded by $e^- e^+$ pairs
  close to the black hole, and by $e^- p$ pairs for $r\gtrsim 10r_g$.
  \item BZ-power and neutron diffusion primarily determines Lorentz factor.
  \item Variability is due to toroidal field instabilities.
  \item Poynting flux is converted into enthalpy flux and leads to the
  formation of a ``magnetic fireball.''
  \item Patchy jet develops $10^2\lesssim \Gamma_\infty\lesssim 10^3$,
  as required by internal shock model.
  \item Random number of patches ($<1000$ for 30 second burst) and so
  random number of pulses.
  \item Energy structure of jet is Gaussian with $\theta_0\approx
  8^\circ$.
  \item Core of jet with $\theta_j\approx 5^\circ$ can explain GRBs.
  \item Extended slower jet component with $\theta_j\approx 25^\circ$
  can explain x-ray flashes.
  \item Coronal outflows with $\Gamma\sim 1.5$ may power supernovae
  (by producing, e.g., $^{56}{\rm Ni}$) with $M\sim 0.1\msun$
  processed by corona.
\end{enumerate}

Based upon prior numerical \citep{mckinney2005b} and this theoretical
work, basic conclusions for AGN or x-ray binaries include:
\begin{enumerate}
  \item Poynting-dominated jets $e^- e^+$ pair-loaded unless advect
  complicated field.
  \item $\gamma$-ray radiative efficiency, and so pair-loading,
  determines maximum possible Lorentz factor.
  \item Poynting-lepton jet is collimated with $\theta_j\approx 5^\circ$.
  \item Extended slow jet component with $\theta_j\lesssim 25^\circ$.
  \item For fixed accretion rate, variability is due to toroidal field
  instabilities.
  \item Poynting flux is shock-converted into enthalpy flux.
  \item In some AGN, shock heat in transonic transition lost to
  synchrotron emission and limits achievable Lorentz factor to
  $2\lesssim \Gamma\lesssim 10$  (e.g. in M87).
  \item Coronal outflows produce broad inner-radial jet features in AGN
  together with well-collimated jet component (e.g. in M87).
  \item In some x-ray binaries, Compton drag loads Poynting-lepton jets and limits
  Poynting-lepton jet to $\Gamma\lesssim 2$ or jet destroyed.
  \item In some x-ray binaries, Poynting-lepton jet optically thick and emits
  self-absorbed synchrotron.
  \item Coronal outflows have collimated edge with $\Gamma\lesssim 1.5$.
  \item Coronal outflows may explain all mildly relativistic and
  nonrelativistic jets in radiatively efficient systems (most x-ray
  binaries).
\end{enumerate}
For AGN and X-ray binaries, the coronal outflow collimation angle is
strongly determined by the disk thickness.  The above assumed $H/R\sim
0.2$ near the black hole and $H/R\sim 0.6$ far from the black hole,
while $H/R\sim 0.9$ (ADAF-like) is perhaps more appropriate for some
systems.  The sensitivity of these results to $H/R$ is left for future
work.

\section*{Acknowledgments}

I thank Avery Broderick for an uncountable number of inspiring
conversations.  I also thank Charles Gammie, Brian Punsly, Amir
Levinson, and Ramesh Narayan, with whom each I have had inspiring
conversations.  This research was supported by NASA-ATP grant
NAG-10780 and an ITC fellowship.

\begin{appendix}

\section{Pair Creation Notes}\label{GRMHD}
The electron-positron pair plasma that forms may annihilate itself
into a fireball if the pair annihilation rate is faster than the
typical rate of the jet ($c^3/GM$) near the black hole.  Also, if the
pair annihilation timescale is shorter than the dynamical time, then
pair annihilation would give a collisional term in the Boltzmann
equation.  From the pair annihilation rate given by
equation~\ref{taupa}, one finds that $t_{pa}\gg GM/c^3$ for AGN and
marginally so for x-ray binaries.  Thus, pairs mostly do not
annihilate, and so formally the pair plasma that forms in the
low-density funnel region is collisionless so that the Boltzmann
equation should be solved directly. Plasma instabilities and
relativistic collisionless shocks are implicitly assumed to keep the
pairs in thermal equilibrium so the fluid approximation remains mostly
valid, as is a good approximation for the solar wind (see,
e.g. \citealt{fm97,usmanov2000}).  This same approximation has to be
invoked for the thick disk state in AGN and x-ray binaries, such as
for the ADAF model \citep{mckinney2004}.  For regions that pair
produce slower than the jet dynamical time, each pair-filled fluid
element has a temperature distribution that gives an equation of state
with $P=\rho_{0,e^- e^+} k_b T_e/m_e$ rather than $P=(11/12)a T^4$,
where $a$ is the radiation constant.  So most of the particles have a
Lorentz factor of $\Gamma_e\sim u/(\rho_{0,e^- e^+}c^2)$ and little of
the internal energy injected is put into radiation.  This also allow
the use of a single-component approximation.  A self-consistent
Boltzmann transport solution is left for future work.

On the contrary for GRB systems, due to the relatively high density of
pairs, the time scale for pair annihilation is $t_{pa}\ll GM/c^3$
along the entire length of the jet.  Thus a pair fireball forms and
the appropriate equation of state is that of an
electron-positron-radiation fireball.  Thus, formally the pair
fireball rest-mass density is not independent of the pair fireball
internal energy density.  However, because the pairs are well-coupled
to the radiation until a much larger radius of $r\sim 10^8-10^{10}
r_L$, the radiation provides an inertial drag on the remaining pair
plasma.  That is, the relativistic fluid energy-momentum equation is
still accurate.  So the effective rest-mass density is $\sim \rho_0+u$
($u$ the total internal energy of the fireball), and so the effective
rest-mass is independent of the cooling of the fireball until the
fireball is optically thin (see, e.g., \citealt{mr97}).

For GRB systems, the mass conservation equation is reasonably
accurate.  Even though the electron-positron pairs annihilate, the
rest-mass of pairs injected is approximately that of the pairs that
are injected due to Fick-diffusion of neutrons (see next section).
The annihilation energy from electron-positron pairs contributes a
negligible additional amount of internal energy, so can be neglected,
especially compared to the Poynting energy flux that emerges from the
black hole.  Thus, the rest-mass can always be assumed to be due to
baryons rather than the electron-positron pairs.  This also suggests
that the neutrino annihilation is a negligible effect if the BZ power
is larger than the neutrino annihilation power.

In summary, the rest-mass evolution discussed in \citet{mckinney2005b}
are accurate for GRB, AGN, and marginally so for x-ray binaries.  This
is despite the lack of Boltzmann transport for the collisional system,
or a collisional term due to pair annihilation.

\subsection{Baryon Contamination}

Notice that some fraction of baryons contaminate the jet due to
neutron diffusion and subsequent collisional cascade into an
electron-proton plasma \citep{le2003}.  They estimate the diffusion
using Fick's law.  First, the role of ambipolar diffusion is
considered (see, e.g. \citealt{shu92}, chpt. 27).  The drift velocity
is
\begin{equation}
v_{drift,pn} \sim \frac{B^2}{4\pi \gamma_{pn} n_p m_p n_n m_n L} ,
\end{equation}
where $L\sim r(H/R)$ is the typical field radius of curvature induced
by disk turbulence and
\begin{equation}
\gamma_{pn} = \frac{\langle\sigma_{pn} v_{rel,pn}\rangle}{m_p + \rho/m_n}
\end{equation}
is the drag coefficient and $\langle \sigma v\rangle \sim
40c\times 10^{-27}$.  The drift velocity can also be written as
\begin{equation}
v_{drift,pn}/c \sim \left(\frac{b^2}{\rho c^2}\right)\left(\frac{m_n c}{ \langle\sigma v\rangle\rho r(H/R)}\right) .
\end{equation}

Assuming all the diffused neutrons are converted to protons+electrons
and carried with the outflow, then the diffusion flux is
\begin{equation}
F = \rho v_{drift,pn} = \frac{b^2}{\rho c^2}\frac{m_n c^2}{
  \langle\sigma v\rangle r (H/R)} .
\end{equation}
and the mass flux across an axisymmetric conical outflow is
\begin{equation}
\dot{M}_{inj,ambi} = 2\pi \int_0^{r_{out}} F r dr
\end{equation}
and so
\begin{equation}
\dot{M}_{inj,ambi} = 2\pi (F r) r_{out}
\end{equation}
GRB numerical GRMHD models show that the coronal region next to the
Poynting-dominated jet has a time-averaged value of $b^2/(\rho
c^2)\sim 1$ and the turbulent induced eddies occur when the disk has
$H/R\sim 0.2$ \citep{mg04,mckinney2004}.  This gives a mass flux
vs. radius of
\begin{equation}
\dot{M}_{inj,ambi} \sim 10^{-14} \left(\frac{r-r_{stag}}{r_g}\right) \dot{M}_{acc} ,
\end{equation}
where pairs that enter inside $r\lesssim r_{stag}$ fall into the black
hole and do not load the jet.  For a recombination radius of $r_n\sim
2\times 10^9$cm \citep{le2003}, this gives
\begin{equation}
\dot{M}_{inj,ambi} \sim 4\times 10^{-10} \dot{M}_{acc} .
\end{equation}

This can be compared to the result of \citet{le2003} for
Fick-diffusion, where the mass injection rate of free-streaming
particles (their eq. 7) is
\begin{equation}
\dot{M}_{inj,Fick} \sim 3\times 10^{-7} \left(\frac{r}{r_g}\right)^{2/3}
\dot{M}_{acc}
\end{equation}
for their collapsar model with jet half-opening angle of
$\theta_j=0.1$, specific enthalpy $h\sim 1$, $L_j\sim 10^{51}\ergps$,
a neutron thermal velocity of $v\sim 0.1c$.  The thermal velocity is
based upon the near-funnel coronal value of $u/(\rho_0 c^2)\sim
0.01-0.1$ as measured from GRMHD numerical models.  Notice that this
ratio is typically $0.01$ in the corona, but is $0.1$ at the edge, so
we use $0.1$ since the Fick diffusion is based upon the edge values.
Notice that they used $v\sim c$.  For a recombination radius of
$r_n\sim 2\times 10^9$cm after which no more free neutrons exist, one
has that
\begin{equation}\label{fickmdot}
\dot{M}_{inj,Fick} \sim 7\times 10^{-5} \dot{M}_{acc} .
\end{equation}
Hence, Fick diffusion dominates ambipolar diffusion.  The role of
reconnection, between the corona/coronal wind and jet, in loading the
jet with baryons is left for future work.

The characteristic timescale for moving these pairs is $\sim t_g
(r/r_g)$ and characteristic length is $\sim r_g (r/r_g)$, so a
characteristic density vs. radius is
\begin{equation}\label{fickrho}
\rho_{p e^-} \sim \frac{\dot{M}_{inj,Fick}t_g}{r_g^3} \left(\frac{r}{r_g}\right)^{-2}
\sim 3\times 10^{-7} \left(\frac{r}{r_g}\right)^{-4/3} \rho_{0,disk} .
\end{equation}
Notice that this is comparable to the rest-mass in pairs given by
equation~\ref{collapsardensity}.  Thus, as the fireball decays in pair
rest-mass, the rest-mass quickly becomes dominated by neutrons
diffusing across the magnetic wall between the corona and funnel.
Hence, the baryon conservation law holds and the approximations used
here hold well. A pair-annihilation term is only needed to account for
the contribution to the internal energy.  Since $f_\rho\sim 8f_h$,
this contribution is a $\sim 10\%$ effect and is not expected to
affect the results of the numerical models of \citet{mckinney2005b}.

\subsection{Pair-Radiation Equation of State}

The total amount of comoving energy put into the thermal fireball is
\begin{eqnarray}\label{u0tot}
u_{0,tot}=\rho_{0,e^- e^+}c^2+u_{0,e^- e^+}+u_{0,\gamma} =\nonumber \\
AT^4\int_0^\infty dx
\frac{x^2\sqrt{x^2+\tilde{m}^2}}{e^{\sqrt{x^2+\tilde{m}^2}}+1} +u_{0,\gamma} ,
\end{eqnarray}
where $u_{0,\gamma}=1.62348 A T^4$, $A=4.66244\times 10^{-15} \erg
\cm^{-3} {\rm K}^{-4}$, $x\equiv pc/k_b T$, $\tilde{m}\equiv m_e
c^2/k_b T$, $p$ is the momentum in the fluid frame, and the
rest-number density of photons is $n_\gamma=20.2944\cm^{-3}T^3$.  The
rest-mass in pairs is
\begin{equation}
\rho_{0,e^- e^+} =
B T^3 \int_0^\infty dx \frac{x^2}{e^{\sqrt{x^2+\tilde{m}^2}}+1} ,
\end{equation}
where $B=3.07589\times 10^{-26}{\rm g} \cm^{-3} {\rm K}^{-3}$, and the
number density of pairs is $n_{0,e^- e^+}=\rho_{0,e^- e^+}/m_e$.
Notice that the GRMHD equations of motion relate the comoving energy
to energy at infinity by
\begin{equation}
u_{tot}=u_{0,tot} u^t u_t+p_{gas}=(f_\rho+f_h+f_m)e_{\nu\bar{\nu},ann}
\end{equation}
and $\rho_{e^- e^+}=\rho_{0,e^- e^+} u^t$, where
\begin{equation}
p_{gas}= \frac{AT^4}{3}\int_0^\infty dx \frac{x^2\sqrt{x^2+\tilde{m}^2}}{e^{\sqrt{x^2+\tilde{m}^2}}+1} + p_{0,\gamma} ,
\end{equation}
and $p_\gamma = u_{0,\gamma}/3$.  This gives sufficient information to
solve for $u^t$ and $u_t$ or $\dot{\rho}_{e^- e^+} =\partial/\partial
t(\rho_{0,e^- e^+} u^t)$.  The study of \citet{mckinney2005b} uses a
fixed $\gamma$-law gas equation of state with $\gamma=4/3$ to model
the typically radiation-dominated system.

\section{Ideal MHD Quantities Conserved along each Flow Line}\label{conservedflow}

Kerr spacetime is stationary and axisymmetric with 2 Killing vectors
$\xi_t^\mu=\frac{\del}{\del_t}=-\delta^\mu_t$ and
$\xi^\phi_\mu=\frac{\del}{\del\phi}=\delta^\mu_\phi$ that satisfy
$\mathcal{L}_\mathbf{\xi}(\mathbf{g})=0$, where $\mathcal{L}$ is the
Lie derivative and $\mathbf{g}$ is the metric.  For a vector $X^\mu$,
tensors $\mathbf{T}$ that obey $\mathcal{L}_X(\mathbf{T})=0$ are
conserved along $X$.  In particular, for $X=\xi$, such a tensor is a
physical quantity independent of the ignorable coordinates $t$ and
$\phi$.  For $X=u$, the 4-velocity, the tensor is conserved along each
flow line.  One can derive a set of conserved flow quantities that are
associated with the Killing symmetries \citep{bek78}.  The below
summarizes those results that are key to this paper.  This
presentation is necessary for the discussion regarding the
determination of the Lorentz factor of the jet.

In the ideal MHD approximation one can show that
$\mathcal{L}_u(\xi^\mu A_\mu)=0$, where $F_{\mu\nu} \equiv A_{\nu,\mu}
- A_{\mu,\nu}$ defines the vector potential $A_{\mu}$.  Thus for an
unsteady axisymmetric flow the $\phi$ component of the magnetic vector
potential ($A_\phi=\xi_\phi^\mu A_\mu$) is conserved along each flow
line, while for a steady non-axisymmetric flow the electric potential
($A_t=\xi_t^\mu A_\mu$) is conserved along each flow line.
\citet{bek78} also show that there are four other independent
conserved scalar quantities for an axisymmetric, stationary fluid.
These correspond to the first integrals of Maxwell's equation
$\dF^{\mu\nu}_{;\mu}=0$ and the conservation equation
$T^\mu_{\nu_;\mu}=0$ with the use of the continuity equation $(\rho_0
u^\mu)_{;\mu}=0$.  These first integrals are associated with the
projection of the 2 Killing vectors and the magnetic field ($b^\mu$)
on these equations of motion.

For a degenerate ($\dF^{\mu\nu} F_{\mu\nu} = e^\mu b_\mu = 0$),
stationary and axisymmetric plasma, one finds
\begin{equation}
A_{\phi,\theta} A_{t,r} - A_{t,\theta} A_{\phi,r} = 0.
\end{equation}
It follows that one may write
\begin{equation}\label{omega}
{A_{t,\theta}\over{A_{\phi,\theta}}} = {A_{t,r}\over{A_{\phi,r}}} \equiv -\Omega_F(r,\theta)
\end{equation}
where $\Omega_F$ is usually interpreted as the ``rotation frequency''
of the electromagnetic field (this is Ferraro's law of isorotation;
see e.g. \citealt{fkr}, \S 9.7 in a nonrelativistic context).  Notice
that $\Omega_F \equiv F_{tr}/F_{r\phi} = F_{t\theta}/F_{\theta\phi}$.
One can show in the ideal MHD approximation that
\begin{equation}\label{OMEGAF}
\Omega_F = v^\phi - B^\phi v^\theta/B^\theta = v^\phi - B^\phi v^r/B^r
\end{equation}
is conserved along each flow line.  The first term corresponds to
fluid rotation and the second term corresponds to the slip along the
toroidal component of a field line.  This yields $F_{\mu\nu}$ in terms
of the free functions $\Omega_F, A_\phi$, and $B^\phi$.  Thus, $F_{rt}
\equiv \detg E_r = -\Omega_F A_{\phi,r}$, $F_{\theta t} \equiv \detg
E_\theta = - \Omega_F A_{\phi,\theta}$, $F_{r\theta} = \detg B^\phi$,
$F_{\phi r} = \detg B^\theta = - A_{\phi,r}$, and $F_{\theta\phi} =
\detg B^r = A_{\phi,\theta}$.  The diagonal components are zero and
$F_{\phi t} \equiv E_\phi=0$, where $B^i \equiv \dF^{it}$, $E_i \equiv
F_{it}/\detg$ such that $B^i E_i =\dF^{\mu\nu}
F_{\mu\nu}/(4\detg)\equiv E\cdot B$.  Thus for fixed poloidal magnetic
field, $\Omega_F$ is a measure of the electric field.  With the
Faraday written in terms of $B^i$ and $\Omega_F$, the electromagnetic
field automatically satisfies the source-free Maxwell equations.

Using similar constraints on $F^{\mu\nu} = A^{\nu,\mu} - A^{\mu,\nu}$
and with $E^i \equiv F^{it}\detg$, $B_i \equiv \dF_{it}$, one can show
that $\detg F^{rt} \equiv E^r = \tau_\theta B_\theta$, $\detg
F^{\theta t} \equiv E^\theta = -\tau_r B_r$, $\detg F^{\phi t} \equiv
E^\phi = \tau_\phi B_\phi$, $\detg F^{r\theta} = -B_\phi$, and $\detg
F^{r\phi} = B_\theta$, $\detg F^{\theta\phi} = - B_r$.  Here there is
only one independent quantity among the three $\tau_i$'s that are set
by $\dF_{\mu\nu} F^{\mu\nu}=0$ and that the flow and metric are
stationary and axisymmetric (i.e. $E_\phi=0$).  One can show that
$\tau_\phi = B_r B_\theta (\tau_r-\tau_\theta)/B_\phi^2$ and solve for
another by using $E_\phi=F^{\alpha\beta} g_{\alpha t} g_{\beta
\phi}=0$.

It is interesting to note that in Boyer-Lindquist coordinates
$\tau_\phi=0$ and $\tau_r=\tau_\theta\rightarrow \tau$ and then the
contravariant and covariant Faraday take on the same simple form with
\begin{equation}
\tau=-\frac{g^{t\phi} - g^{tt}\Omega_F}{g^{\phi\phi} -
g^{t\phi}\Omega_F}=\frac{\Omega_F/\Omega_{ZAMO} - 1}{\Omega_0 -
\Omega_F},
\end{equation}
where $\Omega_{ZAMO} \equiv g^{t\phi}/g^{tt}=2ar/A$ is the angular
frequency of a zero angular momentum observer (ZAMO) and
\begin{equation}
\Omega_0 \equiv g^{\phi\phi}/g^{t\phi} =
\frac{a^2-\Delta/\sin^2{\theta}}{2ar}.
\end{equation}
Notice that if $\Omega_F=\Omega_{ZAMO}$, then $\tau=0$ and so $E^i=0$.
So the difference between the dragging of inertial frames and the
field rotation frequency generates the electric field $E^i$.  Also in
Boyer-Lindquist $B_r = B^r |g|
g^{\theta\theta}g^{\phi\phi}(\Omega_F/\Omega_0 - 1)$, where $|g|
g^{\theta\theta} g^{\phi\phi} = (\Sigma-2r)/\Delta$.  Also, $B_\theta
= B^\theta |g| g^{rr}g^{\phi\phi}(\Omega_F/\Omega_0 - 1)$, where $|g|
g^{rr} g^{\phi\phi} = \Sigma-2r$.  Also, $B_\phi = -B^\phi |g| g^{rr}
g^{\theta\theta}$, where $|g| \equiv |{\rm Det}(g_{\mu\nu})|$ and $|g|
g^{rr} g^{\theta\theta} = \Delta\sin^2{\theta}$.

Using the definition of $F^{\mu\nu}$ given above one can define
$A_\phi$ in terms of the poloidal $B^{r}$ and $B^{\theta}$ giving
\begin{equation}\label{APHIEQ}
A_{\phi}(l_f)-A_{\phi}(l_i) = \int_{l_i}^{l_f} (\detg (B^r d\theta - B^\theta dr))
\end{equation}
over the line segment from $l_i$ to $l_f$.  Thus given the poloidal
field components, then $A_\phi$ can be determined up to a constant.
The contours of constant $A_\phi$ represent time-dependent poloidal
magnetic field surfaces for any $\phi$.  Shown in Cartesian
coordinates, beyond a few gravitational radii from even a $j=1$ black
hole, the density of lines represents the field strength in the lab
frame.  Near the horizon where the intrinsic volume of space is larger
than in Minkowski space-time, the density of field lines in such a
Cartesian plot overestimates the lab frame field strength by factors
of $\lesssim 2$.

For an inviscid fluid flow of magnetized plasma, the energy and
angular momentum flux per unit rest-mass flux
\begin{equation}\label{ECONS0}
E = -T^r_t/(\rho_0 u^r)=-T^\theta_t/(\rho_0 u^\theta)
\end{equation}
and
\begin{equation}\label{LCONS0}
L = T^r_\phi/(\rho_0 u^r)=T^\theta_\phi/(\rho_0 u^\theta) ,
\end{equation}
respectively, are conserved along each flow line.  For unmagnetized
flows $E$ is conserved for any stationary flow, while $L$ is conserved
for any axisymmetric flow.  If the ideal MHD approximation ($e^\mu =
0$) holds, then the magnetic flux per unit rest-mass flux
\begin{equation}\label{PHI}
\Phi = \frac{B^r}{\rho_0 u^r} = \frac{B^\theta}{\rho_0 u^\theta} = \frac{B^\phi}{\rho_0(u^\phi-u^t \Omega_F)}
\end{equation}
is conserved along each flow line.  This also implies that $(\Phi +
b^t)/\rho_0 = b^r/v^r = b^\theta/v^\theta$.  \citet{bek78} also show
that
\begin{equation}\label{PSI}
\Psi=-E+\Omega_F L = h(u_t + \Omega_F u_\phi) ,
\end{equation}
is conserved along each flow line.  One can reduce $E$ and $L$ to
forms such as
\begin{equation}\label{ECONS}
E = -\zeta u_t - \Psi\Phi b_t/h = -h u_t + \Phi\Omega_F B_\phi
\end{equation}
\begin{equation}\label{LCONS}
L = \zeta u_\phi + \Psi\Phi b_\phi/h = h u_\phi + \Phi B_\phi
\end{equation}
where $h=(\rho_0+\IEDEN+p)/\rho_0$ is the gas specific enthalpy and
$\zeta = h+b^2/\rho_0$ is the total specific enthalpy.  Notice that
for an isentropic flow that $d(\rho_0+u)/d\rho_0 = h$ and so
\begin{equation}\label{isentropic1}
dp/(h\rho_0) = dh/h .
\end{equation}
Clearly any ratio of these conserved flow quantities is also conserved
along each flow line (e.g. the energy flux per unit magnetic flux
($E/\Phi$)).  Any axisymmetric, stationary flow solution can be
written in terms of the 6 independent quantities $A_\phi$, $B^\phi$,
$\Omega_F$, $\Phi$, $E$, and $L$, where the single function $A_\phi$
determines the dependent quantities $B^r$ and $B^\theta$.  Entropy is
a dependently conserved quantity when one writes the rest-mass density
and enthalpy in terms of the entropy and another conserved
quantity. In general, the solution is determined once the conserved
flow quantities are set by the boundary conditions and the other
quantities are set by the equations of motion, either directly or
using the Grad-Shafranov approach (for a review see, e.g.,
\citealt{lev05}).  The limitations of, and hence extensions to, the
ideal MHD approximation are described in \citet{meier2004}.

Note that for an axisymmetric stationary degenerate fluid the
Boyer-Lindquist coordinate components are related to the Kerr-Schild
coordinates by $B^r[\BL] = B^r[\KS]$ , $B^\theta[\BL] =
B^\theta[\KS]$,$B^\phi[\BL] = B^\phi[\KS] - B^r[\KS] (a - 2
r\Omega_F)/\Delta$, while for a general fluid $B_r[\BL] = B_r[\KS] +
(a B_\phi)/\Delta$, $B_\theta[\BL] = B_\theta[\KS]$, and $B_\phi[\BL] = B_\phi[\KS]$.

\section{Ideal MHD Fluid Forces}\label{fluidforces}

First, to investigate the spatial acceleration of the fluid along a
flow line $A_\phi$ one requires a unit length space-like vector that
satisfies $V^\mu A_{\phi,\mu}=0$ since $A_{\phi}$ is conserved along a
flow line.  The unit-length magnetic field satisfies these properties
since $B^\mu A_{\phi,\mu}=0$.  To study the poloidal acceleration
along a field line, the toroidal magnetic field component it projected
out to obtain the unit vector
\begin{equation}
B^{\hat{\mu}}_p = N B^\mu_p = N (B^\mu - \frac{(\omega^\phi_\nu
  B^\nu)\xi_\phi^\mu}{|\omega^\phi||\xi_\phi|}) = N(B^\mu -
  B^\phi\delta^\mu_\phi)
\end{equation}
where $N=1/\sqrt{B^\mu_p B^\nu_p g_{\mu\nu}} = 1/\sqrt{B^i B^j
g_{ij}}$ with $\{i,j\}=\{r,\theta\}$ only, and $\bf{\omega}^\phi$ is the
$\phi$ basis one-form.  Therefore the projection of an acceleration
along a poloidal projection of each flow line is
\begin{equation}\label{ACCAPHI}
a_{A_{\phi}} = a_\mu B^{\hat{\mu}}_p = a_r B^{\hat{r}} + a_\theta B^{\hat{\theta}}
\end{equation}

Second, to investigate the spatial collimation of the fluid, a unit
length space-like vector that is perpendicular to the poloidal field
line and perpendicular to the $\phi$-direction (i.e. $\xi_\phi$) is
required.  This vector is
\begin{equation}
C^{\hat{\mu}} = N_C C^\mu = N_C \epsilon^\mu_{\alpha\beta}
B^{\hat{\alpha}}_p \xi^\beta_\phi ,
\end{equation}
where $N_C = 1/\sqrt{C^\mu C^\nu g_{\mu\nu}}$ and
${\epsilon^\mu}_{\alpha\beta}$ is the spatial permutation tensor.  Thus
$C^{\hat{r}} = N N_C B^\theta$ and $C^{\hat{\theta}} = N N_C B^r$.
Therefore the projection of an acceleration in the collimation
direction is
\begin{equation}\label{ACCCOLL}
a_{coll} = a_\mu C^{\hat{\mu}} = N_C (a_r B^{\hat{\theta}} - a_\theta
B^{\hat{r}}) .
\end{equation}
Notice that for $\theta<\pi/2$ if $a_{coll}<0$, then the flow is
collimating toward the polar axis.  For $\theta>\pi/2$ if
$a_{coll}>0$, then the flow is collimating toward the polar axis.

Third, there are many interesting frames to measure the acceleration.
The acceleration away from geodesics is obtained from the projection
of $P^{\mu\nu} = g^{\mu\nu} + u^\mu u^\nu$ on $\nabla_\gamma
T^\gamma_\mu=0$, giving Euler's equations for the deviation from
geodesic motion
\begin{equation}\label{force0}
\rho_0 h a^G_\mu = -{P_\mu}^\alpha p_{,\alpha} + J^\alpha F_{\alpha\mu}
\end{equation}
where $a_G^\mu=u^\mu_{;\nu} u^\nu =
u^\alpha(u^\mu_{,\alpha}+{\Gamma^\mu}_{\beta\alpha} u^\beta)$ is the
``geodesic acceleration'' away from the geodesic motion ($a_G^\mu=0$).
Note that $\Gamma$ here is the connection coefficient and not the
Lorentz factor.  This comoving geodesic acceleration ``hides'' the
effect of gravity on the fluid.  One could instead focus on the
coordinate acceleration $a_C^\mu=u^\nu u^\mu_{,\nu}$, which represents
the change in the 4-velocity in the momentarily comoving frame.  From
the geodesic equation of motion,
\begin{equation}\label{COORDACC1}
a_C^\mu = a_G^\mu - {\Gamma^{\mu}}_{\alpha\beta} u^\alpha u^\beta ,
\end{equation}
and so the coordinate acceleration along a flow line is
\begin{equation}\label{COORDACC2}
a^C_{A_\phi} = a^C_\mu B_p^{\hat{\mu}} =  a^G_{A_\phi} - B^{\hat{\mu}}
u^\alpha u^\beta \Gamma_{\mu\alpha\beta},
\end{equation}
so the geodesic deviation and gravitational acceleration along each
flow line can be studied separately.

Notice that for a stationary flow $a^C_{A_\phi}=0$ where the poloidal
velocity $u^p=0$.  For a black hole with field lines that tie the
black hole to large radii, there must exist a region where $u^p=0$.
For an outflow to reach large distances, the region where $u^p=0$
separates those fluid elements that eventually fall into the black
hole and those fluid elements that reach large distances.
\citet{lev05} refers to this as the ``injection surface.''  If
particles were created only due to reaching the Goldreich-Julian
charge density, then this must be the location where particles emerge.
For an injection region with negligible angular velocity
$u^\phi\approx 0$, then
\begin{equation}\label{COORDACC3}
a^G_{A_\phi} \approx  B^{\hat{r}} \Gamma_{r tt}+B^{\hat{\theta}} \Gamma_{\theta tt}
\end{equation}
determines the location of the stagnation surface.  In Boyer-Lindquist
coordinates
\begin{equation}
\Gamma_{r tt} = \frac{r^2 - (a\cos{\theta})^2}{\Sigma^2}
\end{equation}
and
\begin{equation}
\Gamma_{\theta tt} = -\frac{2ra^2\cos{\theta}\sin{\theta}}{\Sigma^2}.
\end{equation}
See section~\ref{nonidealmhd} for a discussion of injection physics.

\subsection{Forces in Lab Frame}

The forces as written in the lab frame, rather than comoving frame,
allow for simple understanding of the force dynamics.
Equations~\ref{force0} and~\ref{ACCAPHI} imply that
\begin{equation}
a^G_{A_\phi} = N B^j ( {a^G_j}^{(MA)} + {a^G_j}^{(EM)} ) ,
\end{equation}
where for a stationary, axisymmetric flow
\begin{equation}\label{ACCMA1}
{a^G_j}^{(MA)} = \frac{-u_j u^i p_{,i} + p_{,j}}{\rho_0 h}
\end{equation}
is the hydrodynamic acceleration.  For an isentropic flow,
equation~\ref{isentropic1} implies that
\begin{equation}\label{ACCMA2}
{a^G_j}^{(MA)} = {P_j}^i (\log{h})_{,i} = -u_j u^i (\log{h})_{,i} + (\log{h})_{,j} .
\end{equation}
For an axisymmetric, stationary, ideal MHD fluid the electromagnetic
acceleration ${a^G_j}^{(EM)} = J^\alpha F_{\alpha j}/(\rho_0 h)$
reduces to simply
\begin{equation}\label{ACCEM1}
{a^G_j}^{(EM)} = \frac{-B^\phi B_{\phi,j}}{\rho_0 h} .
\end{equation}
Thus the magnetic acceleration is due to the gradient of the toroidal
magnetic field along a field line.  Notice that from
equation~\ref{ECONS} or~\ref{LCONS} that in the limit $\rho_0
h\rightarrow 0$ that $B_\phi\rightarrow E/(\Phi\Omega_F) = L/\Phi$, a
conserved flow quantity, thus $B^j B_{\phi,j}=0$ implying the
electromagnetic field is force-free.  However, notice that the
acceleration ${a^G_j}^{(EM)}$ then becomes undefined.

Equations~\ref{force0} and~\ref{ACCCOLL} imply that
\begin{equation}
a^G_{coll} = N N_C B^j ( {a^G_k}^{(MA)} + {a^G_k}^{(EM)} ) {\epsilon^k}_j ,
\end{equation}
where ${\epsilon^k}_j$ is the poloidal permutation tensor and
${a^G_k}^{(MA)}$ is given in equations~\ref{ACCMA1} or~\ref{ACCMA2}.
This hydrodynamic collimation is due to the pressure acceleration in
the comoving frame along a field line but directed to collimate the
flow.  For an axisymmetric, stationary, ideal MHD fluid the
electromagnetic collimation acceleration ($C^{\hat{j}} J^\alpha
F_{\alpha j}/(\rho_0 h) = C^{\hat{j}} {a^G_j}^{(EM)}$) reduces to
\begin{equation}\label{ACCCOLL2}
{a^G_{coll}}^{(EM)} = \frac{N N_C}{\rho_0 h}\left(
\epsilon^{ab} (B^p)^2 f[B_a]
+ {\epsilon_a}^b B^a B^\phi B_{\phi,b}\right) ,
\end{equation}
where $f[B_a]\equiv \left(B_{a,b} - \Omega_F (\tau_a B_a)_{,b}\right)$
and $(B^p)^2 \equiv (B^r)^2 + (B^\theta)^2$ and ${\epsilon_a}^b$ is
the poloidal permutation tensor.  The last term on the right hand side
of equation~\ref{ACCCOLL2} represents the ``hoop-stress'' that leads
to collimation for nonrelativistic winds.  The first two terms on the
right hand side correspond, respectively, to the forces due to
poloidal magnetic stresses and the electric field ($E^i$)
gradients. The latter can collimate relativistic outflows.

\section{Compton Scattering}\label{compton}

In the lab frame, seed photons are Compton upscattered if the energy
of the photon $E_{seed}\ll \Gamma_e m_e c^2$ for an electron Lorentz
$\Gamma_e$.  The upscattering continues until the lab frame photon
energy exceeds electron energy.  In the lab frame, each scatter gives
the photon a new energy $E_{scat}\approx 4\Gamma_e^2 E_{seed}$ if
$E_{seed}<m_e c^2/\Gamma_e$ and $E_{scat}\approx \Gamma_e m_e c^2$
otherwise.  Two limiting scenarios are if the photon crosses the jet
or if the photon travels parallel and within the jet.

The optical depth to Compton scattering for a photon in the rest frame
of the jet electrons is
\begin{equation}
\tau = \int_{l'} \left(\frac{\rho_{0,e^- e^+}}{m_e}\right) \sigma_T dl' ,
\end{equation}
which in the lab frame gives
\begin{equation}
\tau = \int_{l} \left(\frac{\rho_{0,e^- e^+}}{m_e}\right) \sigma_T \Gamma_e(1-\beta\cos{\theta}) dl ,
\end{equation}
where $\beta=v/c$, $\Gamma_e=(1-\beta^2)^{-1/2}$, and $\theta=\pi/2$
corresponds to perpendicular interactions and $\theta=0$ to parallel.
Across the jet
\begin{equation}\label{tauperp}
\tau_{\perp} = \int_{-\theta_j}^{\theta_j} \left(\frac{\rho_{0,e^+
e^-}}{m_e}\right) \sigma_T \Gamma_e r d\theta ,
\end{equation}
while for along the jet
\begin{equation}
\tau_{\parallel} =\int_{r_0}^{\infty} \left(\frac{\rho_{0,e^- e^+}}{m_e}\right) \sigma_T \Gamma_e
(1-\beta) dr ,
\end{equation}
where $\beta=\left(\frac{\Gamma_e^2-1}{\Gamma_e^2}\right)^{1/2}$.  For
$\Gamma_e\gg 1$, $\Gamma_e (1-\beta)\approx (2\Gamma_e)^{-1}$, so
\begin{equation}\label{taupar}
\tau_{\parallel} \approx \int_{r_0}^{\infty}
\left(\frac{\rho_{0,e^+
    e^-}}{m_e}\right)\left(\frac{\sigma_T}{2\Gamma_e}\right) dr ,
\end{equation}
where $\sigma_T$ is the Thomson scattering cross-section and $r_0$ is
some fiducial starting radius in the jet (see
\citealt{rl79,longair92}).

Similar calculations can be used to estimate the forward or backwards
pair-production or pair-annihilation optical depths.  In the lab
frame, for a photon gas moving with $\Gamma$ with one photon energy
$E=\Gamma E'$ and another $>(\Gamma m_e c^2)^2/E$, then the photons
annihilate with a cross section at $E\lesssim 0.5$MeV of
$\sigma\approx \sigma_T$ and the Klein-Nishina corrections for photons
with $E\gtrsim 0.5$MeV modify the cross section such that
$\sigma\propto E^{-1}$.  For a spectrum (number per unit time per unit
area per unit energy) $f E^{-\alpha}$, then for $\alpha=2$, the
average cross section is $\sigma\approx 0.06 \sigma_T$.  In this case
the relevant proper density is $n_\gamma$ when for a photon beam,
where $n_\gamma$ is the number density of (typically fewer) high
energy photons.  See also \citet{ls01}.

For a beam of electrons with velocity $\beta$, $\sigma\approx
\sigma_T/\beta$ for nonrelativistic electrons and for $\Gamma\gg 1$
\begin{equation}
\sigma\approx \frac{3\sigma_T}{8\Gamma}(\log{2\Gamma}-1) .
\end{equation}
The pair annihilation rate is
\begin{equation}\label{taupa}
t^{-1}_{pa} \approx \langle\sigma v\rangle \left(\frac{\rho_{0,e^- e^+}}{m_e}\right) ,
\end{equation}
which can be compared to some dynamical time to determine if pair
annihilation is important.

\end{appendix}

\label{lastpage}

\end{document}